\documentclass[twocolumn,showpacs,preprintnumbers,amsmath,amssymb,latexsym,prd,nofootinbib,floatfix]{revtex4-2}

\usepackage{amssymb,amsfonts,amsmath,slashed}
\usepackage{graphicx}
\usepackage{mathrsfs}

\usepackage[pdftex, pdftitle={Phase diagram of a generalized Stephanov
  model for finite-density QCD}]{hyperref}

\usepackage{color}

\newcommand{\f}[2]{\frac{#1}{#2}}
\newcommand{\tf}[2]{{\textstyle\f{#1}{#2}}}

\newcommand{\la}{\langle}

\newcommand{\ra}{\rangle}

\newcommand{\de}{\partial}

\renewcommand{\Re}{{\rm Re}\,}
\renewcommand{\Im}{{\rm Im}\,}

\newcommand{\tr}{{\rm tr}\,}
\newcommand{\trf}{{\rm tr}\,}
\newcommand{\tra}{{\rm tr}}

\newcommand{\Uc}{{\cal U}}

\newcommand{\act}{\mathscr{S}}
\newcommand{\tsum}{{\textstyle\sum}}

\newcommand{\actim}{\Phi}
\newcommand{\effact}{S_{\mathrm{eff}}}
\newcommand{\ur}{v}
\newcommand{\ui}{v_I}

\begin{document}

\title{Phase diagram of a generalized Stephanov model for
  finite-density QCD}

\author{Gy{\"o}rgy Baranka}
\email{barankagy@caesar.elte.hu}
\affiliation{ELTE E\"otv\"os Lor\'and University, Institute for
  Theoretical Physics, P\'azm\'any P\'eter s\'et\'any 1/A, H-1117,
  Budapest, Hungary}

\author{Matteo Giordano}
\email{giordano@bodri.elte.hu}
\affiliation{ELTE E\"otv\"os Lor\'and University, Institute for
  Theoretical Physics, P\'azm\'any P\'eter s\'et\'any 1/A, H-1117,
  Budapest, Hungary}

\begin{abstract}
  We solve a random matrix model for QCD at finite chemical potential,
  obtained by generalizing the Stephanov model by modifying the
  random-matrix integration measure with a one-parameter trace
  deformation.  This allows one to check how important the integration
  measure is for the qualitative features of random matrix models, as
  well as to test the robustness and universality of the qualitative
  picture of the original model. While for a small trace deformation
  the phase diagram is identical to that of the Stephanov model, for a
  large deformation an exotic phase with spontaneous
  charge-conjugation breaking appears.
\end{abstract}

\maketitle

\section{Introduction} 
\label{sec:intro}

Understanding strongly interacting matter at finite baryon density
from first principles is a formidable problem, due to the well-known
complex action problem of the path-integral representation of the
Quantum Chromodynamics (QCD) partition function at finite
baryochemical potential. In this representation, the weight of a gauge
configuration (after fermions have been integrated out) is generally a
complex number, and while the full partition function is real and
positive, these properties are realized only through large
cancellations among the various contributions. This makes
importance-sampling numerical techniques of limited use, and even an
intuitive understanding is far from being trivial.

Progress in the development of sophisticated numerical methods to
solve, or at least bypass or ameliorate the sign problem in lattice QCD
has been constant but slow, and currently available methods are still
far from being able to attack the physically interesting regimes (see
Refs.~\cite{Guenther:2022wcr,Aarts:2023vsf,Pasztor:2024dpv} for recent
reviews). Lacking an exact solution of the sign problem by a suitable
change of variables, or a sufficiently powerful numerical method to
bring it under control, investigations of finite-density QCD in a
regime where the sign problem is strong have to resort to effective
descriptions of the system. These can be obtained by means of
QCD-inspired microscopic models, such as the
Nambu--Jona-Lasinio~\cite{Torres-Rincon:2017zbr,Braun:2017srn} or
quark meson models~\cite{Kovacs:2016juc}, including approaches based
on the analogy between a dense system of quarks and a superconducting
material at large chemical potential~\cite{Alford:1997zt} (see
Ref.~\cite{Casalbuoni:2006rs} for a review of older results).

A more general, but also less detailed approach is based on the
formulation of random matrix models mirroring the symmetries of QCD
(see
Refs.~\cite{Verbaarschot:2000dy,Akemann:2007rf,Verbaarschot:2009jz}
for a review). These are built by replacing the Dirac operator, that
describes the interaction of quarks with the non-Abelian gauge fields,
with suitable random matrices preserving some of its properties; and
by replacing the integration over gauge configurations with an
integral over these matrices, with a conveniently chosen measure. In
this approach one trades off the control over the microscopic details
of the system for a simpler analytic structure, that often allows for
an exact solution of the model. While on the one hand one can hardly
hope to reproduce all the details of QCD, on the other hand the
generality of this approach, as well as the universality of certain
properties of random matrix models, can lead to useful insights on the
qualitative properties of the system.

The use of random matrix models in QCD has by now a long history.  In
the specific case of nonzero quark chemical potential $\mu$, a variety
of models has been proposed, capturing a varying amount of the general
properties of the system, and succeeding in some cases at providing a
qualitative picture in agreement with expectations from microscopic
effective models~\cite{Stephanov:1996ki,Janik:1996va,Halasz:1996jg,
  Halasz:1997he,Feinberg:1997dk,Halasz:1998qr,Akemann:2002ym,
  Klein:2003fy,Akemann:2003wg,Osborn:2004rf,Akemann:2004dr,
  Akemann:2007rf}. In particular, at zero temperature, the Stephanov
model~\cite{Stephanov:1996ki} displays a first-order phase transition
between a chirally broken and a chirally symmetric phase, although it
has an unphysical dependence on $\mu$ in the chirally broken
phase. This can be fixed in the Osborn model~\cite{Osborn:2004rf},
which also displays a first order phase transition, at the price of
making the integration measure $\mu$ dependent. In the Akemann
model~\cite{Akemann:2002ym}, instead, the system is always in the
chirally broken phase~\cite{Akemann:2003wg}. Reference
\cite{Halasz:1998qr} extends the Stephanov
model~\cite{Stephanov:1996ki} and the Jackson--Verbaarschot model for
QCD at finite temperature and $\mu=0$~\cite{Jackson:1995nf} (see also
Refs.~\cite{Wettig:1995fg,Jackson:1996xt}) to the whole $(\mu,T)$
plane, obtaining a phase diagram with a second-order (at zero mass) or
crossover (at finite mass) line connecting to the critical temperature
at zero chemical potential, joined by a tricritical point (resp.\ a
critical endpoint) to a first-order line reaching the critical
chemical potential at zero temperature. This is in qualitative
agreement with the model calculations mentioned
above~\cite{Torres-Rincon:2017zbr,Braun:2017srn,
  Kovacs:2016juc,Alford:1997zt,Casalbuoni:2006rs}. A further extension
to finite isospin chemical potential was studied in
Ref.~\cite{Klein:2003fy}. In addition to their use as phenomenological
models, random matrix models are employed to better understand the
technical aspects of the sign problem at finite
density~\cite{Han:2008xj,Bloch:2017sex,Giordano:2023ppk}.

One of the aspects of random matrix models that has received less
attention is the choice of integration measure, most often taken to be
the simplest Gaussian measure (see, however,
Refs.~\cite{Vanderheyden:2000ti,Vanderheyden:2011iq}). The general
expectation is that this is not as important for the qualitative
features of the model as the structural properties of the random
matrix replacing the Dirac operator. This is motivated by universality
results concerning spectral correlations, both on
macroscopic~\cite{Ambjorn:1992xu} and
microscopic~\cite{Kanzieper:1998ti} scales, that are model independent
in wide classes of models (see Ref.~\cite{Akemann:2002ch} for the
specific case of non-Hermitian random matrices relevant to
finite-density QCD, and for a list of references). Nonetheless, the
impact of the choice of measure on the phase diagram of the model can
be nontrivial, as different phases of the system can display similar
spectral correlations -- precisely because of their universal
nature. This is especially so in the presence of a sign problem, where
the effects of cancellations between contributions are difficult to
predict. This kind of study may be helpful in formulating
phenomenologically more realistic models.

The main purpose of this paper is to contribute to fill this gap. We
study a random matrix model for QCD based on the Stephanov model of
Ref.~\cite{Stephanov:1996ki}, differing in the use of a
nonconventional integration measure. While in
Ref.~\cite{Stephanov:1996ki} this is the standard Gaussian measure for
complex matrices, here we include also a squared trace term with a
variable coefficient (notice that this type of deformation was not
considered in Refs.~\cite{Ambjorn:1992xu,Kanzieper:1998ti,
  Akemann:2002ch}). We discuss this model both at real and imaginary
chemical potential. In the imaginary case, the Stephanov model is
formally identical to the Jackson--Verbaarschot model of
Ref.~\cite{Jackson:1995nf}, if one identifies the imaginary chemical
potential of the former with the temperature of the latter. We find
that the qualitative features of the models of
Refs.~\cite{Jackson:1995nf,Stephanov:1996ki} are universal to some
extent, but there are cases where a richer phase diagram emerges, with
two separate phase transitions and a chirally symmetric phase in
between, where charge-conjugation symmetry is spontaneously broken.

The plan of the paper is the following. In Sec.~\ref{sec:rmm} we
formulate the model and discuss its general features. In
Sec.~\ref{sec:sp} we solve the model in the large-$N$ limit of random
matrices of infinite size through standard saddle-point techniques. In
Sec.~\ref{sec:solut} we discuss the phase diagram of the model in
detail. In Sec.~\ref{sec:concl} we draw our conclusions. Most of the
technical details are discussed in Appendices~\ref{sec:app_gi} to
\ref{sec:app_mzero_im}.

\section{Random matrix model}
\label{sec:rmm}

Our model is defined by the following partition function,
\begin{equation}
  \label{eq:Z}
  \begin{aligned}
    Z &= \int d^2W \, e^{-N\left(\tr W W^\dag - \f{c}{N}|\tr
        W|^2\right)}   \left(\det \mathcal{M}\right)^{N_f}\,,\\
    \mathcal{M}&\equiv\begin{pmatrix} m & iW + \mu \\ iW^\dag + \mu &
      m
    \end{pmatrix}\,.
  \end{aligned}
\end{equation}
Here $\mathcal{M}$ has the same structure as the Dirac operator in a
gauge field background, with $m$ and $\mu$ representing respectively
the fermion mass and the corresponding chemical potential; $N_f$ is
the number of fermion ``flavors'' in the system, and $W$ is a complex
$N\times N$ matrix, with corresponding integration measure
$d^2W = \prod_{ij}d \Re W_{ij} d\Im W_{ij}$.  The exponential
Boltzmann factor consists of a standard Gaussian weight modified by
the inclusion of a trace-deformation term, controlled by a parameter
$c$, restricted to $c<1$ to make the integral convergent. The model of
Ref.~\cite{Stephanov:1996ki} is obtained by setting $c=0$; the
finite-temperature, zero-density model (in the unitary class) of
Ref.~\cite{Jackson:1995nf} is obtained by setting $c=0$ and
$\mu=i\pi T$ with real $T$.

For real $\mu$ the partition function is real, since the matrix
$\mathcal{M}=\mathcal{M}[W;m,\mu]$ obeys
$\mathcal{M}[W;m,\mu]^* = \mathcal{M}[-W^*;m,\mu]$ and
$\mathcal{M}[W;m,\mu]^\dag = \mathcal{M}[-W;m,\mu]$, and since
changing integration variables to $W \to -W^*$ or $W\to -W$ does not
modify the Boltzmann factor.  We will refer to this property as
charge-conjugation ($C$) symmetry. Moreover, $\mathcal{M}$ satisfies
the relation
$\gamma_5\mathcal{M}[W;m,\mu]\gamma_5 = -\mathcal{M}[W;-m,\mu]$, where
$\gamma_5$ is the block-diagonal matrix
\begin{equation}
  \label{eq:gamma5}
  \gamma_5=  \begin{pmatrix}
    \mathbf{1} & \mathbf{0} \\ \mathbf{0} & -\mathbf{1}
  \end{pmatrix}\,.
\end{equation}
This reflects the chiral symmetry of QCD with massless quarks,
independently of $\mu$ and $c$. Notice that since $\mathcal{M}$ is a
$2N\times 2N$ matrix, at finite $m$ one has
$\det \mathcal{M}[W;m,\mu] = (-1)^{2N N_f}\det \mathcal{M}[W;-m,\mu]
=\det \mathcal{M}[W;-m,\mu] $, and so $Z(-m,\mu,c) =Z(m,\mu,c)$.  On
the other hand the chiral condensate,
\begin{equation}
  \label{eq:ssb_chi}
  \Sigma(m,\mu,c) \equiv \f{\de \log Z(m,\mu,c)}{\de m}\,,
\end{equation}
obeys $\Sigma(-m,\mu,c)=-\Sigma(m,\mu,c)$, and a nonzero value in the
massless limit signals spontaneous breaking of chiral symmetry.

The relation above can be equivalently written as
$\gamma_5\mathcal{M}[W;m,\mu]\gamma_5=\mathcal{M}[-W,m,-\mu]$,
implying $Z(m,-\mu,c)=Z(m,\mu,c)$.  This also implies that $Z$ is real
for a purely imaginary chemical potential $\mu=i\mu_I$. Finally, the
system is symmetric under the transformation
 \begin{equation}
   \label{eq:discr_sym}
   \mathcal{M}\to \gamma_0\mathcal{M}\gamma_0, \qquad W\to
   W^\dag\,,
\end{equation}
where
\begin{equation}
  \label{eq:discr_sym2}
  \gamma_0=
  \begin{pmatrix}
    \mathbf{0} & \mathbf{1}\\
    \mathbf{1}& \mathbf{0}
  \end{pmatrix}\,,
\end{equation}
which can be identified with a sort of charge conjugation.

The technical motivation behind the additional trace term is the
following. At $m=\mu=c=0$, the integrand is invariant under the chiral
unitary transformation $W\to UWV^\dag$, $U,V\in\mathrm{U}(N)$. This
extended chiral symmetry seems rather accidental, without any strong
physical motivation, and already at $c=0$ it is broken at $\mu\neq 0$
even in the chiral limit. While it helps greatly in the study of
spectral correlations when present~\cite{Osborn:2004rf}, there seems
to be no reason besides practical convenience to exclude terms that
break it explicitly from the integration measure. Notice that at any
value of $m,\mu,c$, the integrand is invariant under the diagonal
subgroup of transformations $W\to UWU^\dag$; for $m=\mu=0$ this is
extended by the $\mathrm{U}(1)$ symmetry under $W\to e^{i\phi}W$.

One could also add a term linear in the traces, i.e.,
$J^* \tr W + J \tr W^\dag$. By a simple change of integration
variables, one can show that adding this term is equivalent to adding
an imaginary component to $\mu$, i.e., $\mu \to \mu -\f{i\Re J}{1-c}$,
as well as including an isospin chemical potential
$\mu_{\mathrm{iso}}=-\f{\Im J}{1-c}$, appearing with opposite signs in
the top right and bottom left blocks of $\mathcal{M}$. The first case
is included in the general discussion below. The second case is beyond
the scope of this paper; we just notice that thanks to
Eq.~\eqref{eq:discr_sym}, $Z$ is symmetric under
$\mu_{\mathrm{iso}}\to -\mu_{\mathrm{iso}}$. Notice that for a complex
$\mu=\mu_R+i\mu_I$ with both $\mu_{R,I}\neq 0$ the partition function
is not guaranteed to be real anymore.

Since the argument of the exponential is still quadratic in the matrix
entries, the solution of this model is obtained by the same standard
procedure used in Refs.~\cite{Jackson:1995nf,Jackson:1995nf,
  Wettig:1995fg,Jackson:1996xt,Stephanov:1996ki,Halasz:1998qr}: after
representing the fermion determinant as a Grassmann integral and
integrating out the $N\times N$ matrix $W$ exactly, one uses a
Hubbard-Stratonovich transformation to recast the partition function
as an integral over an auxiliary complex $N_f\times N_f$ matrix $a$
and an auxiliary complex variable $\omega$. Details are reported in
Appendix \ref{sec:app_gi}; here we only report the final result,
\begin{equation}
  \label{eq:action_Grass2}
  \begin{aligned}
    Z &= C\int d^2a \int d^2\omega\, e^{-N S(a,\omega)
    }\,,\\
    S(a,\omega)& \equiv \tra \left(a a^\dag\right) + |\omega|^2-\log\det M\,,\\
    M&\equiv \begin{pmatrix}
      a + m & \mu - \sqrt{\f{f(c)}{N}}  \omega\\
      \mu + \sqrt{\f{f(c)}{N}}\omega^* & a^\dag + m
\end{pmatrix}\,,\\
f(c) &\equiv \f{c}{1-c}\,,
      \end{aligned}
\end{equation}
where the factor $C=(\pi/N)^{N-N_f-1}/(1-c)$ is irrelevant for
thermodynamics as long as $1-c$ is not exponential in $N$.  Here
$d^2a = \prod_{ij}d\Re a_{ij} d\Im a_{ij}$ and
$d^2\omega = d\Re\omega d\Im\omega$.  Notice that in the allowed
domain $c\in (-\infty,1)$, the function $f(c)$ takes values in the
range $f\in (-1,\infty)$. For definiteness, we set the branch cut of
the square root and of the logarithm on the negative real axis.

The only difference between Eq.~\eqref{eq:action_Grass2} and the
analogous expressions in Refs.~\cite{Jackson:1995nf,Stephanov:1996ki}
is the additional integration over $\omega$ in the presence of
$\omega$-dependent terms in the matrix action $S$ and in the
off-diagonal blocks of $M$. The partition function can then be
estimated in the large-$N$ limit in the saddle-point approximation in
a similar fashion.  This requires one to understand first if and how the
additional terms affect the large-$N$ limit. It is clear that the
additional term is of any relevance only if $f(c)=O(N^\beta)$ with
$\beta\ge 1$; otherwise it can be neglected in the large-$N$ limit,
and one recovers the same expressions as in
Refs.~\cite{Jackson:1995nf,Stephanov:1996ki}. If $\beta > 1$, the
fermionic determinant is dominated by the additional term,
$\det M = N^{-1}f(c)|\omega|^2 + o(N^{\beta-1})$, so
\begin{equation}
  \label{eq:seff0}
  S = \tra \left(a a^\dag\right) + |\omega|^2- \log|\omega|^2 -\log
  \f{f(c)}{N} + \ldots\,, 
\end{equation}
where the omitted terms are subleading in $N$. The minimum of $S$ is
at $a=0$ and $|\omega|^2=1$, independently of the parameters of the
model. The free energy, i.e., $S$ evaluated at the saddle point,
diverges logarithmically in $N$, but its derivatives with respect to
$m$ and $\mu$ are finite and identically zero, independently of $c$.

The only interesting case yet to be solved is $\beta=1$.  Notice that
since it is bounded from below, $f(c)$ must take positive values at
large $N$ if it is to be $O(N)$.  Since we are interested in the
large-$N$ limit, we can set $f(c)=\kappa^2 N$ with
$\kappa\in \mathbb{R}$ without loss of generality, finding
$c= \kappa^2 N/(1+ \kappa^2 N)$, and so
\begin{equation}
  \label{eq:seff1}
  \begin{aligned}
    M &= \begin{pmatrix}
      a  + m & \mu -  \kappa\omega\\
      \mu + \kappa \omega^* & a^\dag + m
    \end{pmatrix}\,,\\
    \det M &= \det\big(
    ( a  + m) ( a^\dag + m) +  \kappa^2|\omega|^2 - \mu^2 \\
    & \phantom{=\det()}+ 2i\mu \kappa\Im\omega \big)\,.
  \end{aligned}
\end{equation}

\section{Saddle-point equations}
\label{sec:sp}

We can now proceed with the saddle-point calculation. The discussion
below follows closely that of Refs.~\cite{Witten:2010cx,
  Cristoforetti:2012su}. One first complexifies the integral by
promoting both the real and the imaginary part of the entries of $a$
and of $\omega$ to complex variables. One then identifies the critical
points of the action, defined by
\begin{equation}
  \label{eq:sp0}
  \f{\de \Re S}{\de a_{ij}}  = \f{\de \Re S}{\de\omega}  = 0\,.
\end{equation}
These in turn define pairs of stable and unstable Lefschetz thimbles,
i.e., submanifolds of $\mathbb{C}^{2(N_f^2+1)}$ where the imaginary
part of the action is constant, and the critical point is an absolute
minimum (resp.\ maximum) of $\Re S$. In general, some of the critical
points will be found on the original ``real'' integration manifold
$\mathcal{C}=\mathbb{R}^{2(N_f^2+1)}\sim \mathbb{C}^{N_f^2+1}$, and
some outside of it. The partition function can be recast exactly as a
sum of integrals over stable Lefschetz thimbles, with only those
thimbles contributing for which the corresponding unstable thimble
intersects the original integration manifold. On these thimbles, at
the critical point $\Re S$ is then larger than on $\mathcal{C}$,
except of course if the critical point lies on $\mathcal{C}$. Then,
critical points outside of $\mathcal{C}$ with $\Re S$ smaller than on
$\mathcal{C}$ do not contribute at all, while those with $\Re S$
larger than on $\mathcal{C}$ give contributions that are exponentially
suppressed in $N$, and can be dropped in the large-$N$ limit since
their number is finite and $N$-independent. This allows us then to
drop the critical points outside of the original integration manifold
altogether as soon as there is at least one on it.

The identification of the critical points is simplified by the
following reasoning. Using a singular value decomposition, we set
$a+m = U h\, \Uc U^\dag$, with $h$ diagonal with real and positive
entries, and $U,\Uc$ unitary. One has
\begin{equation}
  \label{eq:sp1}
  \begin{aligned}
    \tra \left(aa^\dag\right) &= \tr h^2 -2m \Re\tra \left(h\Uc
    \right) + N_f m^2\,,
  \end{aligned}
\end{equation}
while $\det M$ depends only on $h^2$, and not on $\Uc$ or $U$.  Since
$\log\det M = \tr\log M$, the minimization problem separates into
$N_f$ independent problems, each involving only one eigenvalue
$h_{ii}$ and the corresponding diagonal entry $\Uc_{ii}$. The minimum
is obviously the same in each case, and since clearly
$\Uc_{ii}=\mathrm{sgn}(m)$ at the minimum, one has
$\Uc = \mathrm{sgn}(m) \mathbf{1}$ and so $a= A \mathbf{1}$ with
$A\in\mathbb{R}$. The $\mathrm{U}(N_f)$ vector flavor symmetry of the
system is therefore unbroken, in agreement with expectations from the
Vafa-Witten theorem for gauge theories~\cite{Vafa:1983tf,
  Aloisio:2000rb,Giordano:2023spj} in spite of the nonpositivity of
the integrand.\footnote{For $m=0$, $\Uc$ does not enter the
  minimization problem and is therefore arbitrary.  However, taking
  the chiral limit $m\to 0$ from real values, the symmetric solution
  $a= A \mathbf{1}$ is selected.}

We can then restrict our problem to minimizing the real part of the
simpler quantity $S( A \mathbf{1},\omega) = N_f \effact(A,\omega)$,
where
\begin{equation}
  \label{eq:sp4}
  \begin{aligned}
    \effact &\equiv A^2 + \f{1}{N_f}|\omega|^2 \\
    &\phantom{=} - \log \left( (A+m)^2 + \kappa^2|\omega|^2 - \mu^2 +
      2i\mu\kappa\Im \omega\right)\,,
  \end{aligned}
\end{equation}
and $\mu=\mu_R + i \mu_I$ is generally complex. Explicitly,
\begin{equation}
  \label{eq:seff_gen}
  \begin{aligned}
    \Re \effact &= A^2 + \f{1}{
      N_f} |\omega|^2    \\
    &\phantom{=} - \f{1}{2}\log \Big( \big((A+m)^2 +
    (\kappa\Re\omega)^2 + (\kappa\Im\omega - \mu_I)^2 \\ &\phantom{=-
      \f{1}{2}\log}- \mu_R^2\big)^2 + 4\mu_R^2(\kappa\Im
    \omega-\mu_I)^2 \Big) \,.
  \end{aligned}
\end{equation}
At $\kappa=0$ the minimum of $\Re \effact $ is at $\omega=0$, and the
minimization problem reduces to that of
Refs.~\cite{Jackson:1995nf,Stephanov:1996ki}. At $\kappa\neq 0$,
writing
$ \kappa^2|\omega|^2= (\kappa\Re\omega)^2 + (\kappa\Im\omega -
\mu_I)^2 + 2\mu_I (\kappa\Im\omega - \mu_I) + \mu_I^2 $, one sees that
at the minimum of $\Re \effact$ one has $\kappa\Re\omega=0$ and
$\mathrm{sgn}(\kappa\Im\omega - \mu_I) = -\mathrm{sgn}\mu_I$.  By a
similar argument, at the minimum $A+m$ and $m$ have the same sign at
$m\neq 0$.\footnote{At $\mu_I=0$ the minimum of $\Re \effact $ is at
  $\Re\omega=0$, while the sign of $\Im\omega$ is
  undetermined. Similarly, at $m=0$ the sign of $A$ is undetermined.
  That $\mathrm{sgn}(A+m)=\mathrm{sgn}(m)$ follows also directly from
  the argument after Eq.~\eqref{eq:sp1}: since $h=C \mathbf{1}$ with
  $C\ge 0$ and $\mathcal{U}=\mathrm{sgn}(m)\mathbf{1}$, one has
  $a+m = C \mathrm{sgn}(m)\mathbf{1} = (A+m)\mathbf{1}$.}  The task
then reduces to minimizing
\begin{equation}
  \label{eq:sp6_0}
  \begin{aligned}
    \mathcal{S} (A,\Omega,m,\mu,\gamma) &\equiv \Re \effact(A,
    i\Omega/\kappa,m,\mu,\kappa) \\ &= A^2 + \f{1}{\gamma}\Omega^2
    -\f{1}{2} \log Q
    \,,\\
    Q &\equiv \left((A+m)^2 + (\Omega-\mu_I)^2 - \mu_R^2\right)^2 \\
    &\phantom{\equiv} + 4\mu_R^2(\Omega-\mu_I)^2 \,,
  \end{aligned}
\end{equation}
with $\Omega\in\mathbb{R}$ and $\gamma \equiv \kappa^2 N_f> 0$.
Correspondingly,
\begin{equation}
  \label{eq:sp7_0}
  \begin{aligned}
    \actim(A,\Omega,m,\mu,\gamma)&\equiv - \Im \effact(A,
    i\Omega/\kappa,m,\mu,\kappa) \\
    &= \arg \Big( (A+m)^2 + (\Omega - \mu_I)^2 \\ &\phantom{= \arg} -
    \mu_R^2 + 2i\mu_R(\kappa\Omega-\mu_I)\Big)\,.
  \end{aligned}
\end{equation}
We now specialize to purely real and purely imaginary chemical
potential.

\subsection{Real chemical potential}
\label{sec:remu}

For $\mu_R=\mu\in\mathbb{R}$, $\mu_I=0$,
\begin{equation}
  \label{eq:sp6}
  \begin{aligned}
    \mathcal{S}(A,\Omega,m,\mu,\gamma) &= A^2 + \f{1}{\gamma}\Omega^2
    -\f{1}{2} \log Q
    \,,\\
    Q &= \left((A+m)^2 + \Omega^2 - \mu^2\right)^2 + 4\mu^2\Omega^2
    \,,
  \end{aligned}
\end{equation}
and 
\begin{equation}
  \label{eq:sp7}
  \begin{aligned}
    \actim(A,\Omega,m,\mu,\gamma)&= \arg \Big( (A+m)^2 + \Omega^2 -
    \mu^2 + 2i\mu\Omega\Big)\,.
  \end{aligned}
\end{equation}
Taking derivatives with respect to $A$ and $\Omega$ and multiplying by
$Q$, one finds the saddle-point equations
\begin{equation}
  \label{eq:sp_re2}
  \begin{aligned}
    0&= AQ -(A+m)\left((A+m)^2 +
        \Omega^2 - \mu^2\right)\,,\\
0&=\f{\Omega}{\gamma}\left(Q - \gamma\left((A+m)^2 +
        \Omega^2 + \mu^2\right)\right)\,.
  \end{aligned}
\end{equation}
Of course, solutions of Eq.~\eqref{eq:sp_re2} leading to $Q=0$ must be
discarded. There are two types of solutions: with $\Omega=0$, and with
$\Omega\neq 0$.  If $\Omega=0$, since $Q=0$ is excluded, one is left
with the cubic equation
\begin{equation}
  \label{eq:sp_re2bis}
  \begin{aligned}
A((A+m)^2  - \mu^2)&= A+m\,.
  \end{aligned}
\end{equation}
This is the same saddle-point equation as in the Stephanov and
Jackson--Verbaarschot models~\cite{Stephanov:1996ki,Jackson:1995nf}.
For $m\neq 0$ and $\mu\neq 0$ its solutions must satisfy $A+m\neq 0$;
for $\mu=0$, $A+m=0$ is a solution of Eq.~\eqref{eq:sp_re2bis}, but it
is not acceptable since it gives $Q=0$. For $m \neq 0$ then the
minimum satisfies $m(A+m)>0$ strictly. Moreover, for $m\neq 0$ the
solutions of Eq.~\eqref{eq:sp_re2bis} must also satisfy
$A\neq 0$.\footnote{If $A+ m=0$ then also $\mu^2 A=0$, and the two
  equations cannot be true at the same time if $m,\mu\neq 0$. If $A=0$
  then also $A+m =0$, and the two equations cannot be true at the same
  time if $m\neq 0$.}  If $\Omega\neq 0$ one finds the system of
equations
\begin{equation}
  \label{eq:sp_re2ter}
  \begin{aligned}
    AQ &=(A+m)\left((A+m)^2 +
      \Omega^2 - \mu^2\right)\,,\\
    Q&= \gamma\left((A+m)^2 + \Omega^2 + \mu^2\right)\,.
  \end{aligned}
\end{equation}
Also in this case solutions must obey $A+ m\neq 0$ if $m\neq 0$, so
for the minimum $m(A+m)> 0$ strictly. However, solutions with $A=0$ do
exist even if $m\neq 0$.\footnote{Since $\Omega\neq 0$ we have
  necessarily $Q\neq 0$, so the first equation in
  Eq.~\eqref{eq:sp_re2ter} implies that if $A+ m=0$ then also $A=0$,
  and the two equations cannot be true at the same time for $m\neq
  0$. On the other hand, for $m\neq 0$ one has that $A=0$,
  $ \Omega = \pm \sqrt{\mu^2-m^2}$ are solutions if $\mu^2>m^2$ and
  $ 2(\mu^2-m^2) = \gamma$.}  We show below that when
$m,\mu,\gamma\neq 0$, this system reduces to a single cubic equation
for $A$.

\subsubsection{Vanishing chemical potential or mass}
\label{sec:gen_mu_m0}

The saddle-point equations simplify considerably at vanishing chemical
potential, $\mu=0$. In this case
$Q=\left((A+m)^2 + \Omega^2\right)^2$, which must be nonzero for the
solution to be viable. One then finds that if $\Omega=0$ then
$A+m\neq 0$, and so the acceptable solutions solve the quadratic equation
\begin{equation}
  \label{eq:case_mu0}
  A(A+m)-1=0\,,
\end{equation}
while if $\Omega\neq 0$ one has
\begin{equation}
  \label{eq:case_mu0_1}
\begin{aligned}
  A(\gamma-1)- m &=0\,,\\
  (A+m)^2 + \Omega^2-\gamma &= 0\,.
    \end{aligned}
\end{equation}
See Appendix \ref{sec:app_muzero_re} for details.  The saddle-point
equations simplify also in the massless case $m=0$. In this case
$Q = \left(A^2 + \Omega^2 - \mu^2\right)^2 + 4\mu^2\Omega^2 =
\left(A^2 + \Omega^2 + \mu^2\right)^2 - 4\mu^2A^2 $, and the
saddle-point equations reduce to
\begin{equation}
  \label{eq:case_m0}
  A(A^2  - \mu^2-1)= 0\,,
\end{equation}
if $\Omega=0$, and to
\begin{equation}
  \label{eq:case_m0_1}
  \begin{aligned}
    A\left((\gamma-1)\left(A^2 +
        \Omega^2 \right) + (\gamma+1) \mu^2\right)&=0\,,\\
    \left(A^2 + \Omega^2 + \mu^2\right) \left(A^2 + \Omega^2 +
      \mu^2-\gamma\right) &= 4\mu^2A^2\,,
\end{aligned}
\end{equation}
if $\Omega\neq 0$. See Appendix \ref{sec:app_mzero_re} for details.

\subsubsection{Nonzero chemical potential and mass}
\label{sec:gen_mu_m_non0}

If both $\mu\neq 0$ and $m\neq 0$ (as well as $\gamma\neq 0$) the case
$\Omega\neq 0$ can be reduced to solving a single cubic equation.
Subtracting $A$ times the second equation from the first one in
Eq.~\eqref{eq:sp_re2ter} and rearranging terms, one finds
\begin{equation}
  \label{eq:sp_re2_extra3}
  \begin{aligned}
    & ((1-\gamma)A+m)\left((A+m)^2 + \Omega^2 \right)\\
    &=\mu^2((1+\gamma) A+m)\,.
  \end{aligned}
\end{equation}
Any $A$ solving this equation cannot make $(1-\gamma)A+m$
vanish,\footnote{If it did, then also $(1+\gamma)A+m$ would have to
  vanish, leading to $A+m=0$ and $\gamma A=0$, which cannot be true at
  the same time.}  and so we find
\begin{equation}
  \label{eq:sp_re2_extra3_bis}
  \begin{aligned}
    \Omega^2 &=-(A+m)^2 + \mu^2\f{(1+\gamma) A+m}{(1-\gamma)A+m}\,.
  \end{aligned}
\end{equation}
Plugging this result back into the second equation in
Eq.~\eqref{eq:sp_re2ter} and rearranging terms, we find
\begin{equation}
  \label{eq:sp_re2_extra4}
  \begin{aligned}
  & 
(A+m)[
    ((1-\gamma)A+m)  (\gamma+
    2(A+m)((1-\gamma)A+m))
\\ &\phantom{(A+m)[]}-2\mu^2(A+m)]=0\,.
\end{aligned}
\end{equation}
Since for $m\neq 0$ one has $m(A+m)>0$, we are left with the cubic
equation
\begin{equation}
  \label{eq:sp_re_extra4_bis}
2\mu^2(A+m)
  =    ((1-\gamma)A+m)(\gamma+
    2(A+m)((1-\gamma)A+m))\,,
\end{equation}
that together with Eq.~\eqref{eq:sp_re2_extra3_bis} fully
characterizes the minimum of the effective action.

\subsubsection{Symmetry breaking}
\label{sec:gen_SB}

Since $\mathcal{S}$ is invariant under $A\to -A$, $m\to -m$, for any
(local or global) minimum $A_0(m_0)$ for $m=m_0$, one has that
$-A_0(m_0)$ is a (local or global) minimum for $m=-m_0$.  This means
that the phase diagram will be symmetric under $m\to -m$.

At $m=0$, $\mathcal{S}$ and $\actim$ depend only on $A^2$, so for any
saddle-point solution with $A=A_0(0)> 0$ there is another solution
with $A=-A_0(0)< 0$, with identical $\mathcal{S}$ and $\actim$ (as
well as the same contribution from quadratic fluctuations around the
saddle point, which we are ignoring here). Since $m(A_0(m)+m)> 0$ at
small but nonzero $m$, one finds
$\lim_{m\to 0^\pm}A_0(m) = \pm |A_0(0)|$, and so spontaneous breaking
of chiral symmetry if $A_0(0)\neq 0$ [see Eq.~\eqref{eq:thermo1}
below].

Since $\mathcal{S}$ is an even function of $\Omega$, while $\actim$ is
an odd function of $\Omega$, for any saddle-point solution
$(A_0,\Omega_0)$ with $\Omega_0\neq 0$ there is another solution
$(A_0,-\Omega_0)$ with identical $\mathcal{S}$ (and quadratic
fluctuations) but opposite $\actim$. This can lead to complications in
defining a free energy for the model in the usual way,
\begin{equation}
  \label{eq:free_en_def}
  \f{F}{N_f}
  \equiv -\lim_{N\to \infty} \f{1}{N N_f}\log Z\,.
\end{equation}
If these solutions are the absolute minima of $\mathcal{S}$, then for
large $N$
\begin{equation}
  \label{eq:sp_re3}
  \begin{aligned}
    Z &\approx e^{-NN_f\mathcal{S}(A_0,\Omega_0)} \left(e^{iN
        N_f\actim(A_0,\Omega_0)} +
      e^{-iNN_f\actim(A_0,\Omega_0)}\right) \\ & =
    2e^{-NN_f\mathcal{S}(A_0,\Omega_0)} \cos(N
    N_f\actim(A_0,\Omega_0))\,.
  \end{aligned}
\end{equation}
The free energy of the model in this symmetric setup is
\begin{equation}
  \label{eq:free_en0}
  \begin{aligned}
    \f{F}{N_f} &= \mathcal{S}(A_0,\Omega_0) -\lim_{N\to \infty}\f{1}{N
      N_f}\log \cos (N N_f\actim(A_0,\Omega_0))\,,
  \end{aligned}
\end{equation}
and it is not clear how to make sense of the limit in the second term
in the general case.\footnote{One certainly has a definite and
  vanishing limit for $\actim = \f{n_1\pi}{2n_2+1}$, for arbitrary
  $n_2$ and $n_1=-2n_2, \ldots, 2n_2+1$. This set is dense in
  $(-\pi,\pi]$, and could be used to define $F$ in the whole interval
  by continuity.  On the other hand, if $N_f$ is odd and
  $\actim=\f{\pi}{2}$, for $N$ odd one finds real Lee-Yang zeros of
  the partition function.}

This problem can be avoided by selecting one of the two solutions by
introducing an infinitesimal sym\-metry-breaking parameter: this is
our approach here. Introducing an infinitesimal $\mu_I$, since in
general $\mathrm{sgn}(\kappa\Im\omega - \mu_I) = -\mathrm{sgn}\mu_I$,
one finds that the solution with $\Omega<0$ (resp.\ $\Omega>0$) is
selected if $\mu_I \to 0^+$ (resp.\ $\mu_I \to 0^-$). Denoting again
by $A_0$, $\Omega_0$ the position of the minimum of $\mathcal{S}$, we
have
\begin{equation}
  \label{eq:sp_re3bis}
  \begin{aligned}
    Z &\approx e^{-NN_f\mathcal{S}(A_0,\Omega_0)} e^{iN
      N_f\actim(A_0,\Omega_0)} \,,
  \end{aligned}
\end{equation}
and
\begin{equation}
  \label{eq:free_en1}
  \begin{aligned}
    \f{F}{N_f} = \mathcal{S}(A_0,\Omega_0) -i \actim(A_0,\Omega_0)
    \equiv \mathcal{F} - i\varphi \,.
  \end{aligned}
\end{equation}
The phase $\varphi$ is generally different from $0,\pm\pi$, so the
free energy has a nonzero imaginary part. This leads to the
spontaneous breaking of $C$ symmetry. If $\Omega=0$, instead, one has
$\varphi =0$ or $\varphi =\pm\pi$, and at $m\neq 0$ one expects a
unique minimum (selecting a unique minimum also as $m\to 0$).
Finally, notice that if at the saddle point $\Omega_0\neq 0$, one
finds using the second equation in Eq.~\eqref{eq:sp_re2ter}
\begin{equation}
  \label{eq:sp_phase_re}
  \begin{aligned}
    \sin \varphi & = \f{2\mu \Omega_0}{\sqrt{Q(A_0,\Omega_0)}} =
    \f{2\mu}{\sqrt{\gamma}}\f{\Omega_0}{\sqrt{ (A_0+m)^2 + \Omega_0^2
        +\mu^2}} \,.
  \end{aligned}
\end{equation}

\subsection{Imaginary chemical potential}
\label{sec:immu}

For $\mu_R=0$, $\mu_I\in\mathbb{R}$,
\begin{equation}
  \label{eq:9}
  \begin{aligned}
    \mathcal{S}_I (A,\Omega,m,\mu_I,\gamma) & \equiv \mathcal{S}
    (A,\Omega,m,i\mu_I,\gamma) \\ & = A^2 + \f{1}{\gamma}\Omega^2 - \log
    Q_I\,,
    \\
    Q_I &= (A+m)^2 + (\Omega-\mu_I)^2\ge 0\,,
  \end{aligned}
\end{equation}
and correspondingly $\Phi = 0$. At the minimum one has
$\mathrm{sgn}(A)=\mathrm{sgn}(m)$ (if $m\neq 0$) and
$\mathrm{sgn}(\Omega)=-\mathrm{sgn}(\mu_I)$ (if $\mu_I\neq 0$), since
this maximizes $Q_I$ at fixed $|A|$ and $|\Omega|$. The saddle-point
equations read as
\begin{equation}
  \label{eq:sp_im2}
  \begin{aligned}
    0 &= AQ_I - (A+m)\,,\\
    0&=    \Omega Q_I - \gamma(\Omega-\mu_I)\,,
  \end{aligned}
\end{equation}
where again solutions leading to $Q_I=0$ must be discarded. These
equations are cubic in $A$ and quadratic in $\Omega$, and cubic in
$\Omega$ and quadratic in $A$, respectively.  Notice that they imply
that $A\neq 0$ if $m\neq 0$, and $\Omega \neq 0$ if $\mu_I\neq 0$ (if
$\gamma\neq 0$).

If $m\neq 0$ and $\mu_I\neq 0$, the solution of the system of
equations Eq.~\eqref{eq:sp_im2} can be reduced to that of a single
quartic equation.  Subtracting $A$ times the second equation from
$\Omega$ times the first one we find
\begin{equation}
  \label{eq:nz_im1}
  \Omega (A(1-\gamma)+m) = -\gamma \mu_I A\,,
\end{equation}
which, since the right-hand side is nonzero, implies $A(1-\gamma)+m\neq
0$, and so
\begin{equation}
  \label{eq:nz_im2}
  \Omega  = -\gamma\mu_I\f{ A}{A(1-\gamma)+m}\,.
\end{equation}
For a solution of Eq.~\eqref{eq:sp_im2}, $A(1-\gamma)+m$ has the same
sign as $A$, and so as $m$; this requirement is nontrivial only if
$\gamma> 1$, in which case it implies $|A|< |m|/(\gamma-1)$. Plugging
Eq.~\eqref{eq:nz_im2} in the first equation in Eq.~\eqref{eq:sp_im2},
multiplying by $ (A(1-\gamma)+m)^2$, and rearranging terms, one
finally finds
\begin{equation}
  \label{eq:nz_im3}
  (A(A+m)-1)    (A(1-\gamma)+m)^2 + \mu_I^2A(A+m)
  =  0\,.
\end{equation}
It is easy to see that the solution must satisfy $0 < A(A+m) <1$.

In this case the minimum $(A_0,\Omega_0)$ is expected to be unique
(except possibly at $m=0$), and so
\begin{equation}
  \label{eq:free_en_im}
  \begin{aligned}
    Z &\approx e^{-NN_f\mathcal{S}_I(A_0,\Omega_0)} \,,&&& \mathcal{F}
    &= \f{F}{N_f} =\mathcal{S}_I(A_0,\Omega_0) \,.
  \end{aligned}
\end{equation}

\subsection{Thermodynamics}
\label{sec:thermo}

We conclude this section by elucidating the connection between the
saddle-point solution and the relevant thermodynamic quantities.  In
the case of real chemical potential, within our approach the free
energy can develop an imaginary part, so we classify transitions in
terms of the analyticity properties of $\Re F = N_f\mathcal{F}$. In
general, the derivative
$\mathcal{F}_x\equiv \f{\de \mathcal{F}}{\de x}$ of $\mathcal{F}$ with
respect to some parameter $x$ is simply
\begin{equation}
  \label{eq:thermo0}
  \begin{aligned}
    \mathcal{F}_x& =\left(\f{\de A_0}{\de x} \f{\de}{\de A} +\f{\de
        \Omega_0}{\de x} \f{\de}{\de \Omega}
      +\f{\de }{\de x}\right)    \mathcal{S}(A,\Omega;x)|_{\substack{A=A_0\\\Omega=\Omega_0}}\\
    &=\f{\de }{\de x} \mathcal{S}(A_0,\Omega_0;x)\,.
  \end{aligned}
\end{equation}
Further using the saddle point equations, we find for the chiral
condensate $\Sigma \equiv \mathcal{F}_m$,
\begin{equation}
  \label{eq:thermo1}
  \begin{aligned}
    \Sigma &= -\f{2}{Q(A_0,\Omega_0)}(A_0+m)\left((A_0+m)^2 +
      \Omega_0^2 - \mu^2\right) \\ &= -2A_0\,,
  \end{aligned}
\end{equation}
while for the quark density
$n\equiv\mathcal{F}_\mu = 2\mu \mathcal{F}_{\mu^2}$,
\begin{equation}
  \label{eq:thermo2}
  \begin{aligned}
    \f{n}{2\mu} &= \f{1}{Q(A_0,\Omega_0)}\left((A_0+m)^2 - \Omega_0^2
      - \mu^2\right) \\ &= \left\{
      \begin{aligned}
        &\f{1}{(A_0+m)^2 - \mu^2}
        \,, &&& \Omega_0 &=0\,,\\
        &\f{1}{\gamma}\f{(A_0+m)^2 -\Omega_0^2 - \mu^2}{(A_0+m)^2
          +\Omega_0^2 + \mu^2}\,, &&& \Omega_0 &\neq 0\,.
      \end{aligned}\right.
  \end{aligned}
\end{equation}
Moreover, plugging our choice for $c$ in Eq.~\eqref{eq:Z}, one finds 
\begin{equation}
  \label{eq:thermo3}
  \begin{aligned}
    \mathcal{F}_\gamma &= - \f{1}{\gamma^2} \lim_{N\to \infty}
    \f{1}{N^2} \Re \left\la \left|\tr W\right|^2\right\ra=
    -\f{\Omega_0^2}{\gamma^2}\,,
  \end{aligned}
\end{equation}
while the derivative of the imaginary part,
$\f{\de \varphi}{\de \gamma}$, is related to the thermodynamic limit
of $\Im \la |\tr W|^2\ra/N^2$. (This generally does not vanish since
we work at infinitesimal but nonzero $\mu_I$ before taking
$N\to\infty$.) We have then
\begin{equation}
  \label{eq:thermo4}
\begin{aligned}
  A_0 &= -\f{\Sigma}{2}\,,\\
  \Omega_0^2 &= \lim_{N\to \infty} \f{1}{N^2}\Re\left\la \left|\tr
      W\right|^2\right\ra \equiv U\,,
\end{aligned}
\end{equation}
as well as the relations
\begin{equation}
  \label{eq:thermo5}
\begin{aligned}
  \left(m-\tf{\Sigma}{2}\right)^2
  &=\mu^2+\tf{2\mu}{n}\,,  &&& \text{if}~U&=0\,, \\
  \f{1-\tf{n\gamma}{2\mu}}{1+\tf{n\gamma}{2\mu}}
  \left(m-\tf{\Sigma}{2}\right)^2 &= \mu^2 + U \,, &&&
  \text{if}~U&\neq 0\,.
\end{aligned}
\end{equation}
For completeness, we report also the derivative with respect to an
imaginary chemical potential at $\mu_I=0$,
\begin{equation}
  \label{eq:sp6_der0}
  \begin{aligned}
\mathcal{F}_{\mu_I}    &= \f{\de}{\de \epsilon} \mathcal{S}
    (A_0,\Omega_0,m,\mu+i\epsilon,\gamma) |_{\epsilon=0} \\
    &=\f{2\Omega_0\left((A_0+m)^2 + \Omega_0^2 + \mu^2\right)
    }{\left((A_0+m)^2 + \Omega_0^2 - \mu^2\right)^2+ 4\mu^2\Omega_0^2} \,.
  \end{aligned}
\end{equation}
For imaginary chemical potential $\varphi=0$, and the free energy is
always real. We have in this case
\begin{equation}
  \label{eq:thermo_im}
  \begin{aligned}
    \Sigma &= -\f{2(A_0+m)}{Q_I(A_0,\Omega_0)} = -2A_0\,,
  \end{aligned}
\end{equation}
while for the quark density 
\begin{equation}
  \label{eq:thermo_im2}
  \begin{aligned}
    n_I(\mu_I) &\equiv i n(i\mu_I)=  \f{\de }{\de \mu_I}
     \mathcal{S}_I(A_0,\Omega_0;m,\mu_I,\gamma) \\ &=
    \f{2(\Omega_0-\mu_I)}{Q_I(A_0,\Omega_0)} =
    2\f{\Omega_0}{\gamma}\,.
  \end{aligned}
\end{equation}
Equation~\eqref{eq:thermo3} still applies, so Eq.~\eqref{eq:thermo4}
is unchanged, and we have the relation
\begin{equation}
  \label{eq:thermo_im3}
\f{n_I^2}{4\gamma^2} =  U \,. 
\end{equation}
In this case $\Im \la \left|\tr W\right|^2\ra$ vanishes identically.

Second derivatives
$\mathcal{F}_{xy} = \f{\de^2\mathcal{F}}{\de x \de y}$ of the
(normalized) free energy are obtained by taking further derivatives of
Eqs.~\eqref{eq:thermo1}--\eqref{eq:thermo3} and
Eqs.~\eqref{eq:sp6_der0}--\eqref{eq:thermo_im2}, obtaining expressions
that involve first derivatives of the saddle-point solution with
respect to $m$, $\mu$, or $\gamma$.  Within a phase, the (real part of
the) free energy is analytic, and relations between the first
derivatives of $A_0$ and $\Omega_0$ (``Maxwell's relations'') are
obtained by imposing continuity of the second derivatives of
$\mathcal{F}$.  Both for real and imaginary chemical potential,
imposing $ \mathcal{F}_{\gamma m} = \mathcal{F}_{m\gamma}$ we find
\begin{equation}
  \label{eq:second_der_re_im}
  \gamma^2  \f{\de A_0}{\de \gamma} =  \Omega_0 \f{\de\Omega_0}{\de m}\,.
\end{equation}
This shows in particular that $A_0$ is independent of $\gamma$ in a
phase where $\Omega_0=0$. Imposing the identity of mixed second
derivatives involving $\mu$ is not particularly illuminating for real
chemical potential, while for imaginary chemical potential one obtains
the simple relations
\begin{equation}
  \label{eq:second_der_imonly}
  -\f{ \de A_0}{ \de \mu_I} = \f{1}{\gamma}\f{\de \Omega_0}{\de m}\,,
  \qquad - \f{\de \Omega_0}{ \de \mu_I} =
  \f{\de \ln \Omega_0} {\de \ln\gamma}\,.
\end{equation}

\section{Phase diagram} 
\label{sec:solut}

In this section we discuss the phase diagram of the model.  For
$\kappa=0$ one gets back the Stephanov model~\cite{Stephanov:1996ki}
for real chemical potential, and the Jackson--Verbaarschot
model~\cite{Jackson:1995nf} for imaginary chemical potential. In this
case $\omega=0$, the relevant saddle-point equations are
Eq.~\eqref{eq:sp_re2bis} and the first equation in
Eq.~\eqref{eq:sp_im2}, in the two cases respectively, and the free
energy is purely real. This is recovered here in the limit
$\gamma\to 0$, which selects the solutions with $\Omega=0$, and so
exactly the same saddle-point equations as in
Refs.~\cite{Jackson:1995nf,Stephanov:1996ki}. We then focus on the
case $\gamma\neq 0$, discussing first the case of vanishing chemical
potential $\mu=0$ at $m\neq 0$, and the massless case $m=0$, followed
by the general case $m,\mu\neq 0$.

\subsection{Case $\mu=0$, $m\neq 0$}
\label{sec:muzero}

\begin{figure}[t!]
  \centering
  \includegraphics[width=0.49\textwidth]{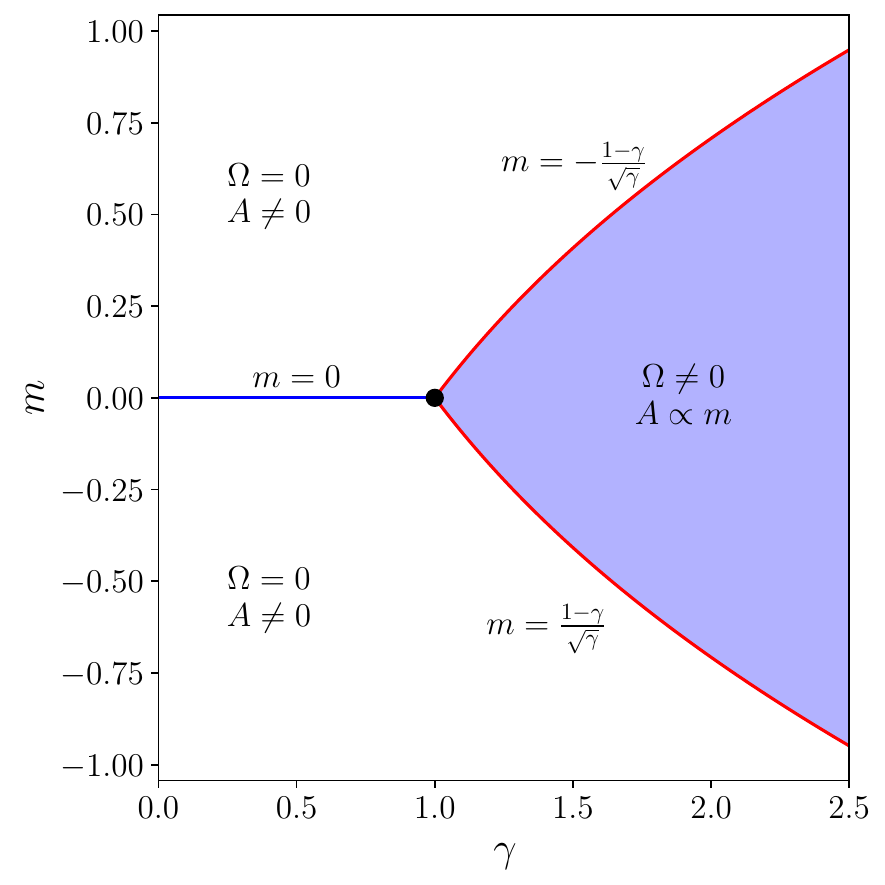}
  \caption{Phase diagram at $\mu=0$ and $m\neq 0$. The blue and red
    lines correspond to first- and second-order transition lines,
    respectively; the blue shaded area is a first-order transition
    surface; the dot marks the junction of the transition lines, where
    the transition is first order.}
  \label{fig:0}
\end{figure}

In this subsection we discuss the case of vanishing chemical
potential, $\mu=0$, in the presence of a nonzero mass, $m\neq 0$. In
this case the saddle-point equations take a very simple form, and a
full analytic solution is straightforward. This is discussed in detail
in Appendix \ref{sec:app_muzero_re}. Set
\begin{equation}
  \label{eq:zeromu_0}
  \begin{aligned}
    \ur(m) &\equiv \f{|m| + \sqrt{m^2+4}}{2}\ge 1\,, \\
    \gamma_0(m) &\equiv \ur(m)^2 = 1+\tf{m^2}{2} + \sqrt{m^2 +
      \tf{m^4}{4}}\,.
  \end{aligned}
\end{equation}
Notice that $1/\ur(m)= \f{-|m| + \sqrt{m^2+4}}{2}$. One distinguishes
three phases, depending on the functional form of the minimum of
$\mathcal{S}$.

\mbox{}

\noindent Phase I: if $m> 0$ and $\gamma < \gamma_0(m)$,
\begin{equation}
  \label{eq:zeromu_1_1}
  \begin{aligned}
    A_0 &=\tf{1}{\ur(m)}\,, &&&
    \Omega_0&=0\,,\\
    \mathcal{F} &= \tf{1}{\ur(m)^2} -\ln \ur(m)^2\,.
  \end{aligned}
\end{equation}
Phase II: if $m< 0$ and $\gamma < \gamma_0(m)$,
\begin{equation}
  \label{eq:zeromu_1_2}
  \begin{aligned}
    A_0 &=-\ur(m)\,, &&&
    \Omega_0&=0\,,\\
    \mathcal{F} &= \tf{1}{\ur(m)^2} -\ln \ur(m)^2\,.
\end{aligned}
\end{equation}
Phase III: if $\gamma > \gamma_0(m)$,
\begin{equation}
  \label{eq:zeromu_1_3}
  \begin{aligned}
    A_0&= \tf{m}{\gamma-1}\,, &&& \Omega_0 & = \pm\sqrt{\gamma -
      \tf{\gamma^2 m^2}{(\gamma-1)^2}}\,,
    \\
    \mathcal{F} &= 1-\tf{m^2}{\gamma-1} -\ln \gamma\,.
  \end{aligned}
\end{equation}
The imaginary part of the free energy vanishes in all three phases,
$\varphi=0$.\footnote{This allows one to define also an ordinary
  thermodynamic limit as in Eq.~\eqref{eq:free_en0}, leading to a
  $C$-symmetric theory at $\mu=0$.}

\mbox{}

In phases I and II one has (loosely speaking, since $m\neq 0$)
spontaneous chiral symmetry breaking with opposite signs of the
condensate, while in phase III one has spontaneous breaking of
charge-conjugation symmetry, with the plus (resp.\ minus) sign chosen
in Eq.~\eqref{eq:zeromu_1_3} if $\mu_I=0$ is approached from negative
(resp.\ positive) values, see Sec.~\ref{sec:gen_SB}.

The phase diagram is shown in Fig.~\ref{fig:0}. There is a line of
first-order phase transitions at $m=0$, where $A_0$ jumps between $1$
and $-1$.  There is also a line of transitions at
$(m,\gamma) = (m,\gamma_0(m))$. Here $\mathcal{F}_m$ and
$\mathcal{F}_\gamma$ are continuous, and $\mathcal{F}_\mu$ is zero
since $\mathcal{S}$ depends on $\mu^2$.  At the point $(m=0,\gamma=1)$
both $\mathcal{F}_m$ and $\mathcal{F}_\gamma$ are discontinuous, as
their limit depends on how this point is approached in the
$(m,\gamma)$ plane. One easily verifies that $\de A_0/\de m$ is
discontinuous on this line of transitions, so $\mathcal{F}_{mm}$ is
discontinuous and the transition is second order. Notice that in the
whole region $\gamma > \gamma_0(m)$ one finds a discontinuous
derivative of both the real and the imaginary part of the free energy
with respect to the imaginary chemical potential since
$|\Omega_0|\neq 0$, the discontinuity vanishing as
$\gamma \to \gamma_0(m)$. This region is then a first-order transition
surface.

\begin{figure}[t]
  \centering
  \includegraphics[width=0.49\textwidth]{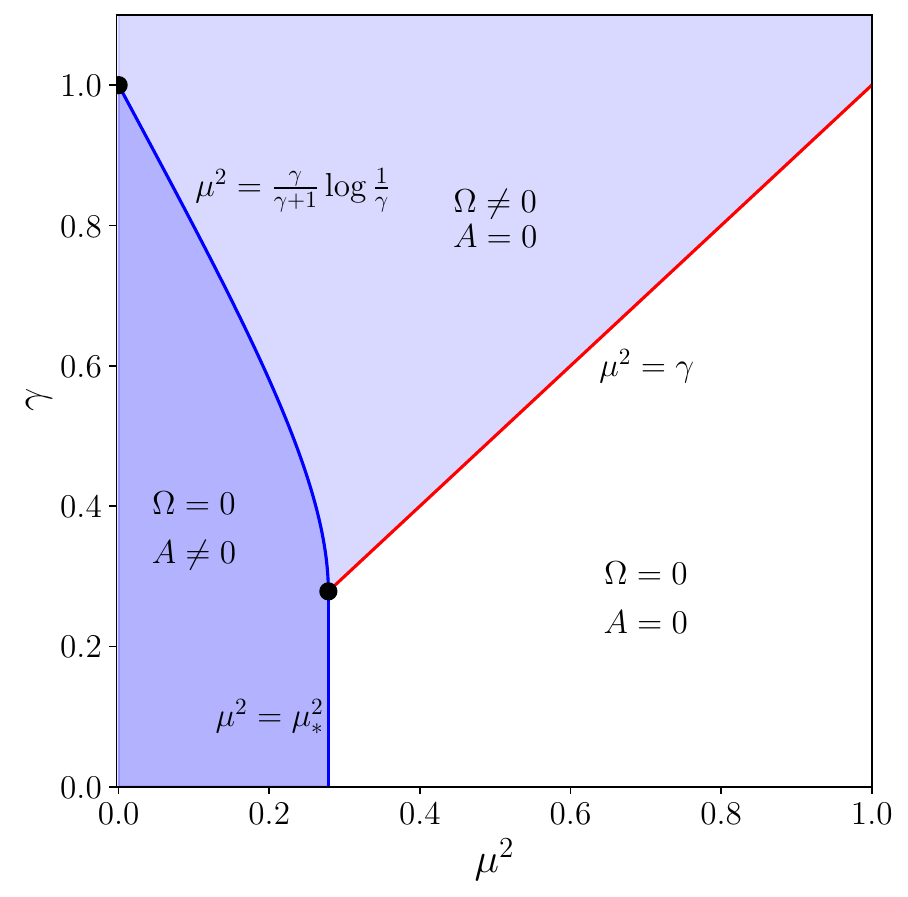}
  \caption{Phase diagram at real $\mu$ and $m=0$. The blue and red
    lines correspond to first- and second-order transition lines,
    respectively; the blue shaded areas are first-order transition
    surfaces; the dot marks the junction of the transition lines,
    where the transition is first order.}
  \label{fig:1}
\end{figure}

\subsection{Case $m=0$}

Also at $m=0$ one can obtain the solutions to the saddle-point
equations in a compact closed form, and the minimum of $\mathcal{S}$
can be identified analytically. We report here the results, leaving
the details for Appendices \ref{sec:app_mzero_re} and
\ref{sec:app_mzero_im}.

\subsubsection{Real chemical potential, $m=0$}
\label{sec:mzero}

Also for real chemical potential at $m=0$ one distinguishes three
phases. Let $\mu_*^2= x_*\simeq 0.2785$ be the solution of
$1+x_* + \ln x_* = 0$, and
$b(\gamma) \equiv -\f{\gamma\ln \gamma}{1+\gamma}$.  The absolute
minimum $(A_0,\Omega_0)$, and the corresponding real and imaginary
parts of the free energy are as follows.

\mbox{}

\noindent Phase I: if $\mu^2\le\min(\mu_*^2,b(\gamma))$,
\begin{equation}
  \label{eq:solu_re_1_1}
  \begin{aligned}
    A_0 &= \pm\sqrt{1+\mu^2} \,, &&& \Omega_0&=0\,,
    \\
    \mathcal{F} &= 1+ \mu^2\,, &&& \varphi &=0\,.
  \end{aligned}
\end{equation}
The positive (resp.\ negative) sign of $A_0$ is selected if $m=0$ is
approached from positive (resp.\ negative) values.

\mbox{}

\noindent Phase II: if $b(\gamma)\le \mu^2 \le \gamma$,
\begin{equation}
  \label{eq:solu_re_1_2}
  \begin{aligned}
    A_0 &= 0 \,, &&& \Omega_0&=\pm\sqrt{\gamma-\mu^2}\,,
    \\
    \mathcal{F}&= 1- \tf{\mu^2}{\gamma} -\ln \gamma\,, &&& \varphi&=
    \pm 2\arcsin\tf{ \mu}{\sqrt{\gamma}}\,.
  \end{aligned}
\end{equation}
Here $\arcsin(x)\in \left[-\tf{\pi}{2},\tf{\pi}{2}\right]$.  The
positive (resp.\ negative) sign of $\Omega_0$ and of $\varphi$ is
selected if $\mu_I=0$ is approached from negative (resp.\ positive)
values.

\mbox{}

\noindent Phase III: if $\mu^2 \ge \max(\mu_*^2,\gamma)$,
\begin{equation}
  \label{eq:solu_re_1_3}
  \begin{aligned}
    A_0 &= 0 \,, &&& \Omega_0 &= 0\,,
    \\
    \mathcal{F}& = -\ln \mu^2\,, &&& \varphi &= \pm\pi\,.
  \end{aligned}
\end{equation}
The sign of $\varphi$ is selected as in phase II.

\mbox{}

The real part of the free energy has discontinuous derivatives
$\mathcal{F}_{m}$ and $\mathcal{F}_{\mu}$ along the line
$L_1=\{\mu^2=\mu_*^2,~ \gamma<\mu_*^2\}$, where $\mathcal{F}_{\gamma}$
is continuous, while all three $\mathcal{F}_{m}$ and
$\mathcal{F}_{\mu}$, $\mathcal{F}_{\gamma}$ are discontinuous on the
line $L_2=\{0<\mu^2<\mu_*^2,~ \mu^2=b(\gamma)\}$. These are then two
lines of first-order transitions. On the line
$L_3=\{\gamma=\mu^2,~\mu^2>\mu_*^2\}$, all three $\mathcal{F}_{m}$,
$\mathcal{F}_{\mu}$, and $\mathcal{F}_{\gamma}$ are continuous, while
$\mathcal{F}_{\mu\mu}$, $\mathcal{F}_{\gamma\gamma}$, and
$\mathcal{F}_{\mu\gamma}$ are not. This is then a line of second-order
transitions. At the point $X_2=(\mu^2=0,\gamma=1)$, $\mathcal{F}_{m}$
and $\mathcal{F}_{\gamma}$ are discontinuous while $\mathcal{F}_{\mu}$
is continuous, and $\mathcal{F}_{\mu\mu}$, $\mathcal{F}_{\mu\gamma}$,
and $\mathcal{F}_{\gamma\gamma}$ are discontinuous.  At the triple
point $X_1=(\mu^2=x_*,\gamma=x_*)$, $\mathcal{F}_{m}$ is
discontinuous, $\mathcal{F}_{\mu}$ is discontinuous if approached from
$L_{1,2}$ but continuous if approached from $L_3$, and $F_\gamma$ is
continuous; $\mathcal{F}_{\mu\mu}$ is discontinuous, while
$\mathcal{F}_{\mu\gamma}$ and $\mathcal{F}_{\gamma\gamma}$ are
continuous if this point is approached from $L_1$, and discontinuous
otherwise.

The imaginary part of the free energy changes discontinuously at the
transition lines $L_{1,2}$ and continuously at the line $L_3$ where,
however, its derivative is discontinuous. The discontinuity on $L_1$,
however, is uninteresting: since $\varphi$ is independent of
$m,\mu,\gamma$ in the phases separated by $L_1$, it can be dropped
entirely without affecting any physical quantity.

The phase diagram is shown in Fig.~\ref{fig:1}. At fixed and low
$\gamma<\mu_*^2$ the phase diagram is identical to that of the
(massless) Stephanov model, with a first-order transition at
$\mu=\mu_*$, independently of $\gamma$, between a low-$\mu$ ``chirally
broken'' phase and a high-$\mu$ ``chirally restored'' phase. The whole
chirally broken region is a first-order surface, across which the sign
of $A$ changes as the sign of $m$ does, so that $\Sigma$ is
discontinuous. The situation changes for $\mu_*^2<\gamma<1$, where one
finds two transitions: a first-order one at $\mu^2=b(\gamma)$, and a
second-order one at $\gamma = \mu^2$. The intermediate phase between
the chirally broken and chirally restored phases is characterized by a
nonvanishing value of $\Omega_0$, which indicates the spontaneous
breaking of charge-conjugation symmetry, with the appearance of a
nontrivial imaginary part of the action. This whole region is again a
first-order surface, across which the sign of $\Omega_0$ changes as
minus that of $\mu_I$, so that $\mathcal{F}_{\mu_I}$ (as well as
$\varphi$) is discontinuous. Finally, for $\gamma>1$ the chirally
broken phase disappears entirely, and only the transition between the
$C$-broken and the chirally restored phases remains.

\begin{figure}[t!]
  \centering
  \includegraphics[width=0.49\textwidth]{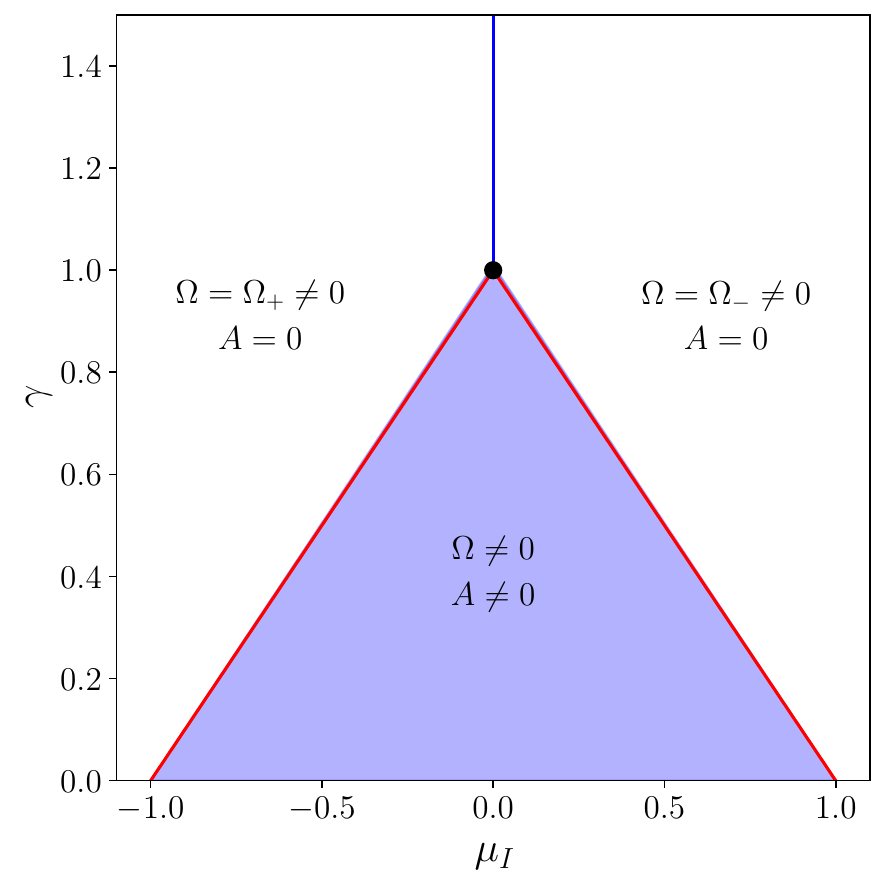}
  \caption{Phase diagram at imaginary $\mu$ and $m=0$. The blue and
    red lines correspond to first- and second-order transition lines,
    respectively; the blue shaded area is a first-order transition
    surface; the dot marks the junction of the transition lines, where
    the transition is first order.}
  \label{fig:2}
\end{figure}

\begin{figure*}[thb!]
  \centering
  \includegraphics[width=0.31\textwidth]{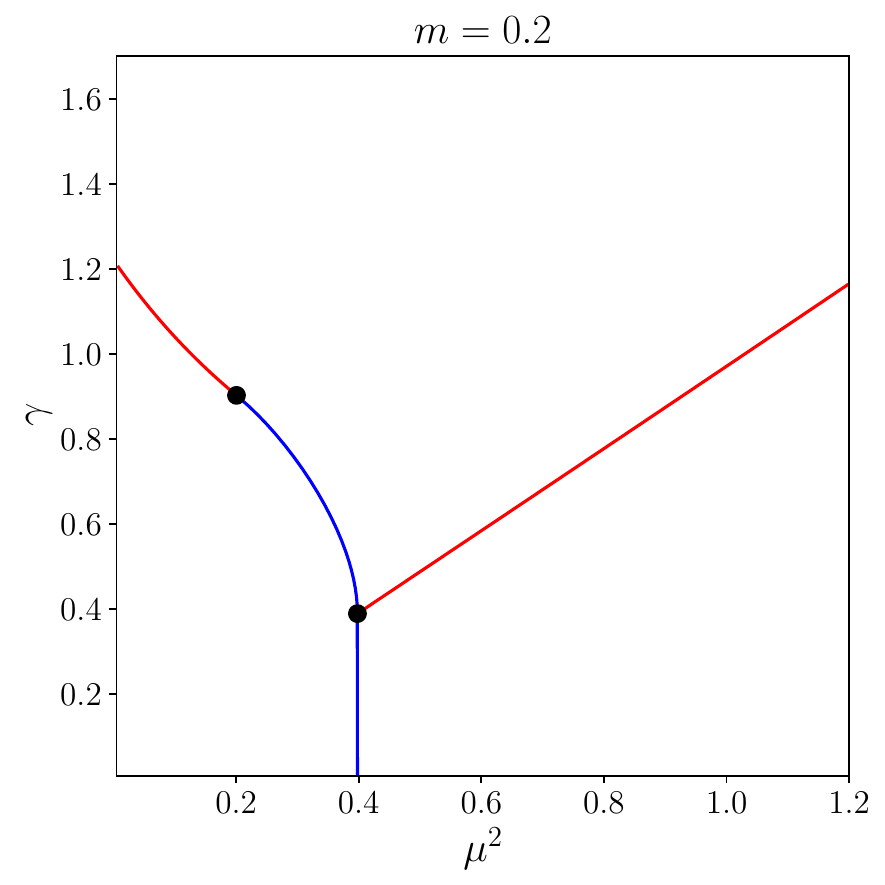}\hfil
  \includegraphics[width=0.31\textwidth]{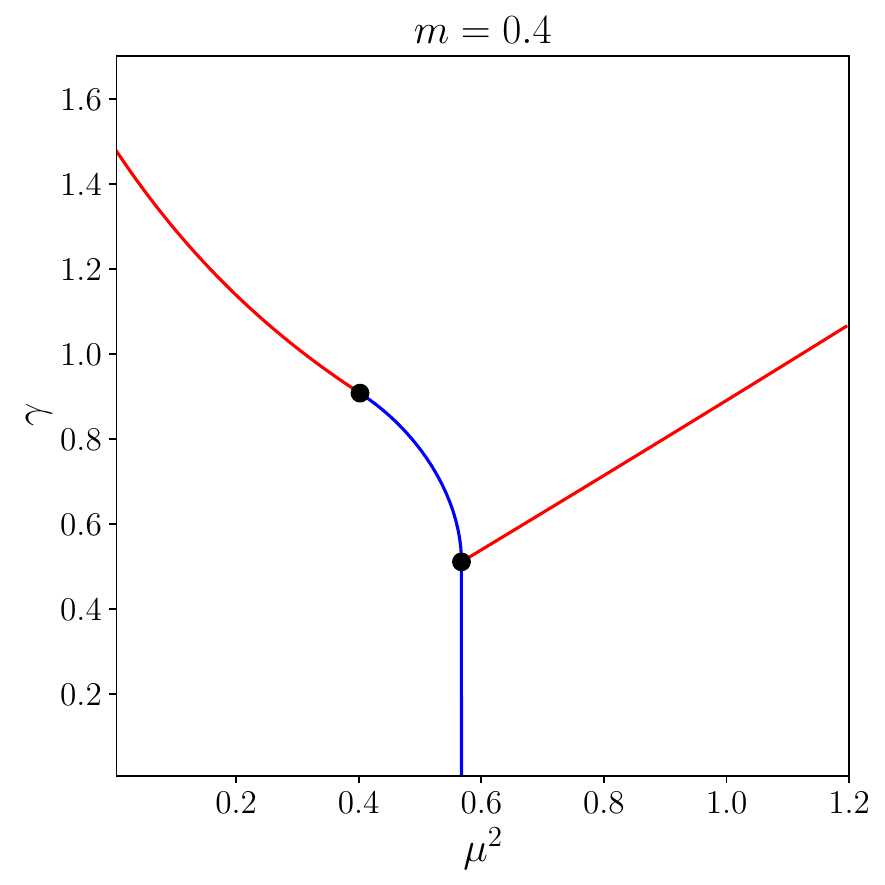}\hfil
  \includegraphics[width=0.31\textwidth]{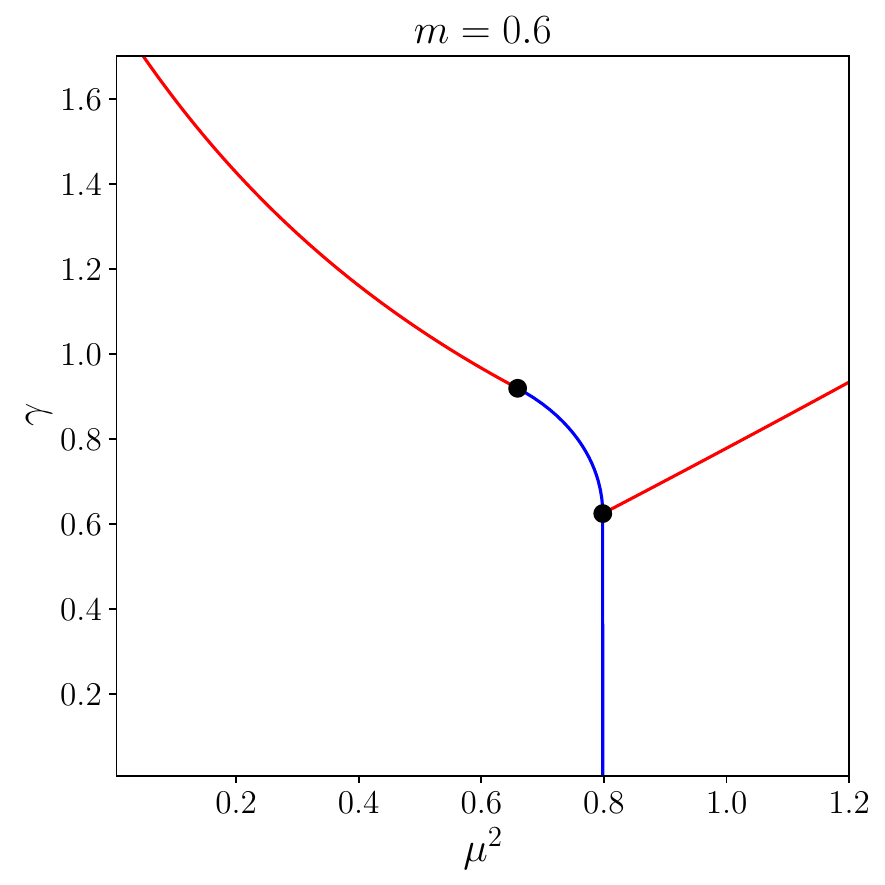}
  \caption{Sections of the phase diagram at real $\mu$ and
    $m=0.2,0.4,0.6$. The blue and red lines correspond to first- and
    second-order transition lines, respectively; dots mark the
    junctions of first- and second-order transition lines.}
  \label{fig:3_0}
\end{figure*}

\begin{figure}[thb]
  \centering
  \includegraphics[width=0.49\textwidth]{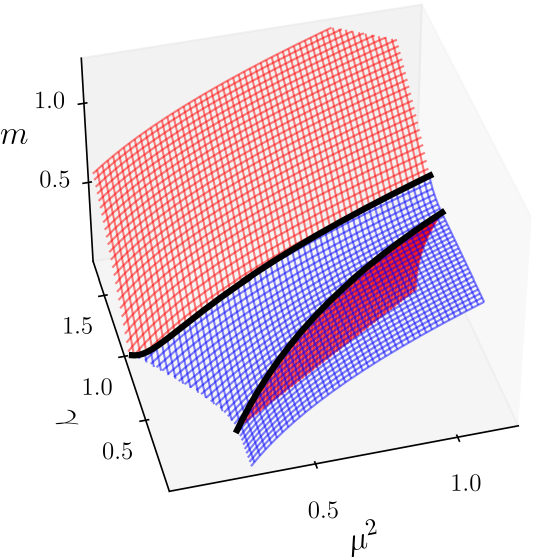}
  \caption{Phase diagram at real $\mu$ and $m\ge 0$.  First-order
    surfaces are shown in blue, second-order ones in red. Black lines
    correspond to their junctions.}
  \label{fig:3}
\end{figure}

\subsubsection{Imaginary chemical potential, $m=0$}
\label{sec:mzeroi}

We now turn to the case of imaginary chemical potential at vanishing
mass. Once again, one distinguishes three phases. Set
\begin{equation}
  \label{eq:solu_im_0}
  \ui (\mu_I) \equiv \tf{|\mu_I| + \sqrt{\mu_I^2 +
      4\gamma}  }{2\sqrt{\gamma}}\ge 1 
  \,.
\end{equation}
Notice that
$1/\ui (\mu_I) = \tf{-|\mu_I| + \sqrt{\mu_I^2 +
    4\gamma}}{2\sqrt{\gamma}}$.

\mbox{}

\noindent Phase I: if $\mu_I\ge \max(1-\gamma,0)$,
\begin{equation}
  \label{eq:solu_im_1_1}
  \begin{aligned}
    A_0&=0\,, \qquad \Omega_0= -\tf{\sqrt{\gamma}}{\ui(\mu_I)} \,,\\
    \mathcal{F}& = \tf{1}{\ui (\mu_I)^2} - \ln\ui(\mu_I)^2 - \ln
    \gamma \,.
  \end{aligned}
\end{equation}
\noindent Phase II: if $\mu_I\le \min(\gamma-1,0)$,
\begin{equation}
  \label{eq:solu_im_1_2}
  \begin{aligned}
    A_0&=0\,, \qquad \Omega_0=   \tf{\sqrt{\gamma}}{\ui(\mu_I)}\,,\\
    \mathcal{F}&=\tf{1}{\ui (\mu_I)^2} - \ln\ui(\mu_I)^2-\ln \gamma\,.
  \end{aligned}
\end{equation}
\noindent Phase III: if $|\mu_I|\le 1-\gamma$,
\begin{equation}
  \label{eq:solu_im_1_3}
  \begin{aligned}
    A_0&= \pm \sqrt{1-\tf{\mu_I^2}{(1-\gamma)^2}}\,, \qquad \Omega_0 =
    -\tf{\gamma\mu_I }{1-\gamma}\,,\\
    \mathcal{F}&= 1 - \tf{\mu_I^2}{1-\gamma} \,.
  \end{aligned}
\end{equation}
The positive (resp.\ negative) sign of $A_0$ is selected if $m=0$ is
approached from positive (resp.\ negative) values.

\mbox{}

The derivative $\mathcal{F}_{\mu_I}$ is continuous on the two
transition lines $L_\pm=\{\mu_I=\pm (1-\gamma),0< \pm\mu_I \le 1\}$,
but discontinuous on the line $L_0=\{\mu_I=0$, $\gamma> 1\}$;
$\mathcal{F}_m$ and $\mathcal{F}_\gamma$ are continuous on all three
transition lines. However, $\mathcal{F}_{\mu_I\mu_I}$ is discontinuous
on $L_\pm$.

The phase diagram is shown in Fig.~\ref{fig:2}. The lines of
second-order transitions $L_\pm$ meet with the first-order line $L_0$
at the triple point $\mu_I=0$, $\gamma=1$. The triangular region where
$A_0\neq 0$ is a first-order surface where $A_0$ changes sign as $m$
does, making the condensate $\Sigma$ discontinuous. In the limit
$\gamma\to 0$ the critical points are found at $\mu_I=\pm 1$, in
agreement with the results of Ref.~\cite{Jackson:1995nf}.

\subsection{Case  $\mu\neq 0$, $m\neq 0$}
\label{sec:mnzero}

One can solve the relevant equations by quadrature also in the general
case $\mu\neq 0$, $m\neq 0$, using Cardano's formula at real chemical
potential, or Ferrari's formula at imaginary chemical potential. Since
the expressions are quite cumbersome we do not report them here. We
also did not attempt an analytic determination of the absolute
minimum, resorting instead to numerical methods. In order to look for
singularities in the derivatives of the partition function, one needs
the derivatives of $A_0$ and $\Omega_0$ with respect to
$x=m,\mu,\gamma$. Since $\de \Omega_0/\de x$ can be expressed in terms
of $A_0$ and $\de A_0/\de x$, one only needs to express
$\de A_0/\de x$ in terms of $A_0$, in order to use the corresponding
numerical solution to evaluate derivatives. This can be done by taking
the derivative of the relevant equation, which yields a linear
equation for $\de A_0/\de x$ that can be solved straightforwardly.

\begin{figure*}[t]
  \centering
  \includegraphics[width=0.31\textwidth]{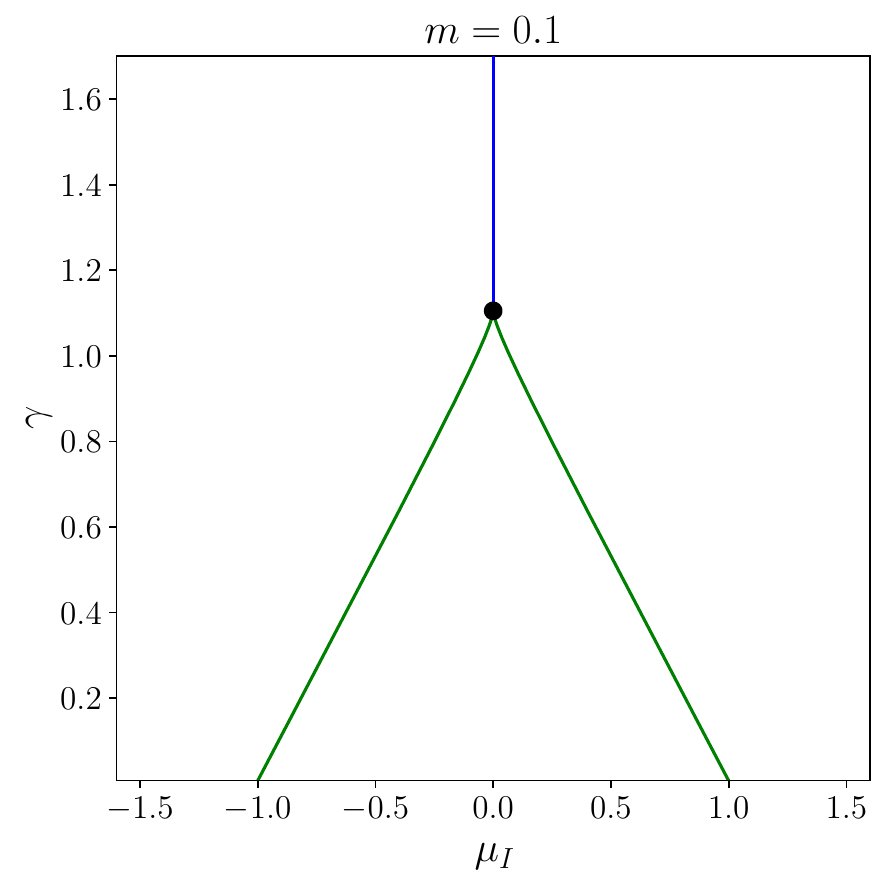}\hfil
  \includegraphics[width=0.31\textwidth]{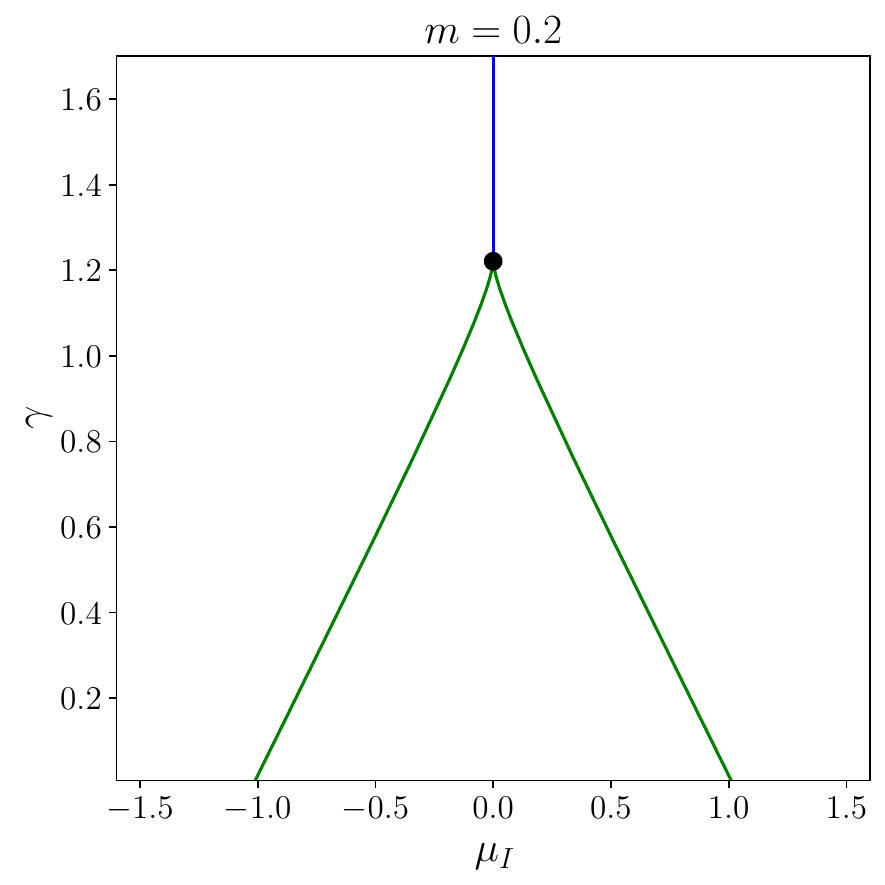}\hfil
  \includegraphics[width=0.31\textwidth]{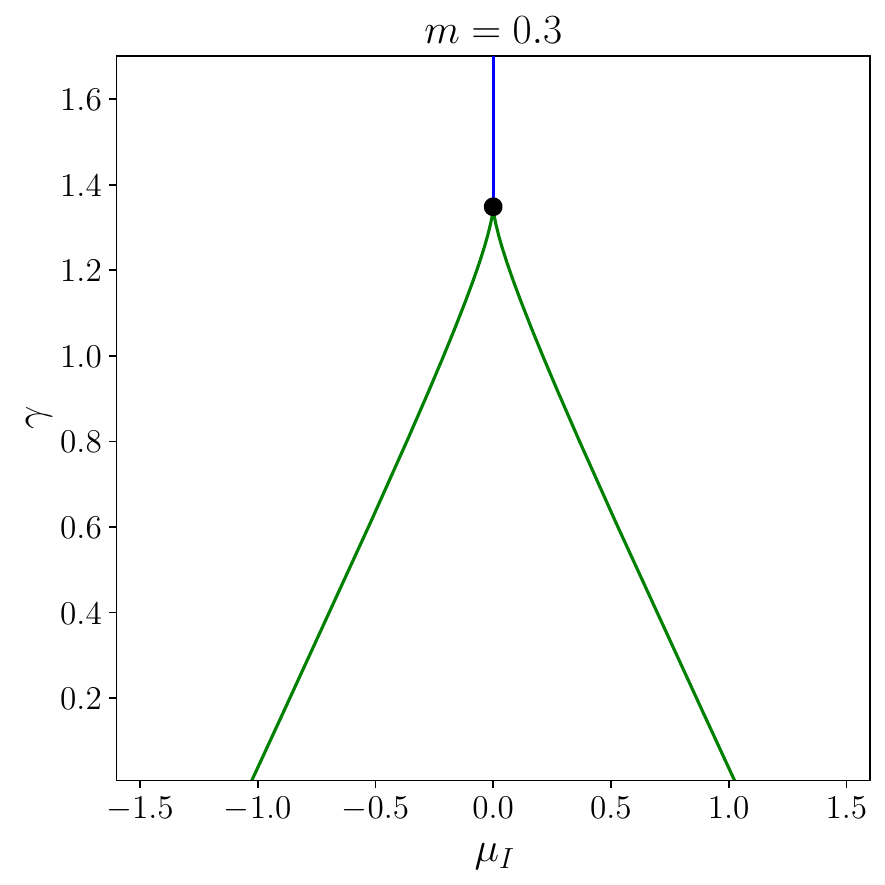}
  \caption{Section of the phase diagram at imaginary $\mu=i\mu_I$ and
    $m=0.1,0.2,0.3$. Blue and green lines correspond to first-order
    transitions and crossovers, respectively; dots mark the junctions
    of crossover and first-order transition lines.}
  \label{fig:4_0}
\end{figure*}
\begin{figure}[thb]
  \centering
  \includegraphics[width=0.49\textwidth]{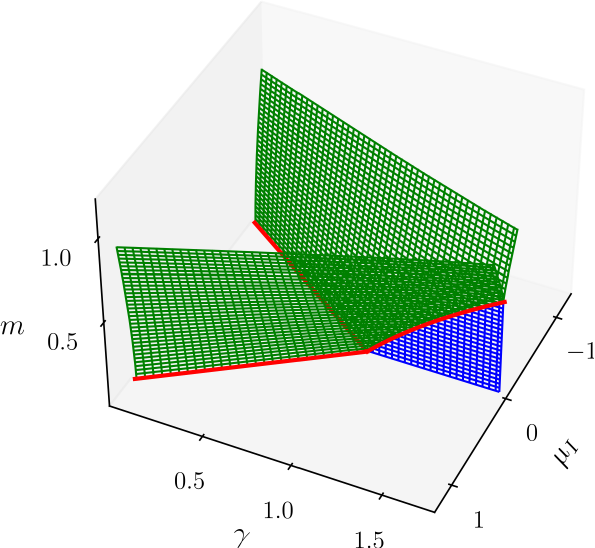}
  \caption{Phase diagram at imaginary $\mu$ and $m\ge 0$.  A
    first-order surface is shown in blue, crossover surfaces in green.
    Red lines correspond to their junction, and to the lines of
    second-order transitions at $m=0$.}
  \label{fig:4}
\end{figure}

\subsubsection{Case $m\neq 0$, real chemical potential}
\label{sec:mnzero_re}

Sections of the phase diagram at fixed $m\neq 0$ and real chemical
potential are shown in Fig.~\ref{fig:3_0}. These are qualitatively
similar to the phase diagram at $m=0$, see Fig.~\ref{fig:1}. In the
lower left corner (phase I) $\Omega=0$, and $A\neq 0$ with $A$ large,
corresponding to spontaneous breaking of chiral symmetry on top of the
explicit breaking due to nonzero $m$. In the lower right corner (phase
II) $\Omega=0$, and $A\neq 0$ but with $A$ small, corresponding to
explicit chiral symmetry breaking effects only. Finally, in the top
part of the diagram (phase III) $\Omega\neq 0$, and $A\neq 0$ with $A$
small: here charge-conjugation symmetry is spontaneously broken. The
transition between phases I and II is first order; since $\Omega=0$
and the relevant saddle-point equation is Eq.~\eqref{eq:sp_re2bis},
the critical chemical potential is $\gamma$ independent, and identical
to that found in Ref.~\cite{Stephanov:1996ki}.  The transition between
phases II and III is instead second order, bending away from a
straight line at large $\mu^2$, while that between I and III has a
second-order and a first-order part. The points separating the first-
and second-order transitions lines are continuously connected to the
points $X_{1,2}$ found at $m=0$. At the transition lines, $\varphi$
displays the same type of singularity as $\Omega$, so it is
discontinuous at a first-order line and has a discontinuous derivative
at a second-order one.

The full phase diagram, including $m=0$, is shown in Fig.~\ref{fig:3}:
a second-order surface joins a first-order one to separate the
chirally broken phase from the rest, which is further split into a
chirally and charge-conjugation symmetric phase and a chirally
symmetric phase where $C$ is spontaneously broken, separated by a
surface of second-order transitions.

\subsubsection{Case $m\neq 0$, imaginary chemical potential}
\label{sec:mnzero_im}

Sections of the phase diagram at $m\neq 0$ and imaginary chemical
potential are shown in Fig.~\ref{fig:4_0}. These are again
qualitatively similar to what is found at $m=0$. Here the minimum is
unique, with $\Omega\neq 0$ everywhere (except at $\mu_I=0$ for
sufficiently small $\gamma$, see below). The top left and top right
regions correspond to chirally restored phases where $A\neq 0$ but
with $A$ small, separated by a line of first-order transitions where
$\Omega$ jumps by $-2\sqrt{\gamma}$, corresponding to the spontaneous
breaking of $C$ if one approaches $\mu=0$ from a nonvanishing
imaginary chemical potential at $\gamma\neq 0$. In the central bottom
region $A\neq 0$ with $A$ large, corresponding to spontaneous chiral
symmetry breaking; here $\Omega=0$ at $\mu_I=0$. However, this region
is separated from the rest only by crossovers. One finds two symmetric
crossover lines (defined as the maxima of $\de A_0/\de \mu_I$),
transformed into each other by $\mu_I\to -\mu_I$, and joining at
$\mu_I=0$, $\gamma= \gamma_0(m)$ (see above Sec.~\ref{sec:muzero}).

The full phase diagram, including $m=0$, is shown in Fig.~\ref{fig:4}:
crossover surfaces, separating the chirally broken phase from the
rest, join a first-order surface at a second-order line. At $m=0$ the
crossovers turn into second-order transitions (see above
Sec.~\ref{sec:mzeroi}).

\section{Conclusions}
\label{sec:concl}

Studies of QCD at finite density using first-principle lattice
calculations are made difficult by a complex-action problem, that
prevents the use of standard and efficient numerical methods. The use
of effective models in elucidating the phase diagram is then a
practical necessity, both to get some physical insight into the problem,
and to better understand how to possibly overcome, or at least
ameliorate, the complex-action problem.

A simple but very general class of models is the random matrix models,
where the details of the gauge field configurations are bundled into a
large random matrix, encoding the interaction between gauge and
fermion fields. Model building is guided here only by the general
symmetry properties of the system, and the details of the matrix
integration measure are not expected to play a major role.

In this paper we have tested this expectation by investigating a
modification of the Stephanov model for finite-density
QCD~\cite{Stephanov:1996ki}, changing only the matrix integration
measure by including a trace deformation in the action. While the
phase diagram is left entirely unchanged by the deformation when this
is not too large, a very different phase diagram is found for a large
deformation, including also a phase where charge-conjugation symmetry
is spontaneously broken.

While there is no particular reason to modify the matrix integration
measure, there is also no particular reason not to, the standard
Gaussian choice being dictated only by simplicity. This measure has a
larger symmetry than the measure employed here. Our results for the
phase diagram lead us to wonder whether this larger symmetry, that
would seem accidental, has instead a deeper meaning.

In a complementary way, we wonder also whether the trace deformation
employed here has some physical meaning, corresponding to some feature
of the gauge fields. Particularly tantalizing is the existence of an
exotic $C$-broken phase in our model, separated by a first-order
transition from the chirally broken phase, of which no analog is known
in QCD. This phase may be related to the possibility of spontaneous
$C$ breaking in QCD and QCD-like theories with one small compactified
dimension if periodic boundary conditions are imposed on fermions (see
Refs.~\cite{Lucini:2007im,Lucini:2009kf}). A much more exciting
possibility is, of course, that it describes an actual phase of QCD at
finite density, with the presence of a first-order transition between
the $C$-broken and the chirally broken phases making it extremely
interesting from a cosmological perspective.  Further speculation
along these lines, however, should wait for a better understanding of
the random-matrix integration measure.

In a more down-to-Earth perspective, it is worth noticing that, in the
interesting cases, in the thermodynamic limit the trace deformation
removes the contribution of the random matrix trace from the action,
making it an effectively flat direction. This is somewhat reminiscent
of the existence of flat (gauge) directions in the configuration space
of gauge fields. In order to make the analogy complete, one should
remove the trace of the random matrix also in the fermionic
determinant, either exactly already at finite $N$, or in an effective
way as with the ``gauge'' part of the action -- in which case the most
natural approach is to use the same coefficient used there. In both
cases, however, one ends up with the trace deformation being entirely
ineffective in the thermodynamic limit, and one finds again the same
phase diagram as in the Stephanov model. Using a (rather contrived)
setup where the removal of the trace is achieved at different rates in
the matrix action and in the fermion determinant in the large-$N$
limit, one reproduces instead the more general phase diagram of
Sec.~\ref{sec:solut} (see Appendix~\ref{sec:app_gi} for details).

At the present stage, it is hard to tell if the highly nontrivial
phase diagram found in the presence of a large trace deformation is a
genuine physical feature of finite-density QCD, or possibly of some
other system with a complex-action problem; Or if it is a consequence
of bad modeling. Either way, our results confirm how difficult it is
to have intuition on systems with a complex-action problem, and how
difficult it is to model them accurately.

From a practical perspective, since our model can be solved exactly,
it provides at the very least a useful benchmark to test methods to
deal with the complex-action problem, also in a very demanding setup
if parameters are tuned to the $C$-broken phase. In particular, since
the effect of the trace deformation is precisely to modify the weight
of the trace of the random matrix in the path integral, it allows one
to test their methods in a setting where one can control the
effectiveness of a simple complex shift of the integration contour in
ameliorating the complex-action problem. Such shifts are known to be
very effective in a variety of models~\cite{Alexandru:2018fqp,
  Tulipant:2022vtk,Rodekamp:2022xpf,Giordano:2023ppk,Gantgen:2023byf},
but they are forbidden in QCD due to the tracelessness of the gauge
field.

From a more optimistic point of view, the universality of the phase
diagram for relatively small deformations supports the qualitative
picture of finite-density QCD obtained from the Stephanov model. This
model is known to have a number of drawbacks (e.g., no ``Silver
Blaze'' phenomenon at low $\mu$~\cite{Cohen:2003kd}, no distinction
between Abelian and non-Abelian gauge fields, no Roberge-Weiss
periodicity at imaginary $\mu$~\cite{Roberge:1986mm}); the fact that
small trace deformations keep the picture intact indicates that they
can reasonably be included in the toolbox when trying to formulate
physically more accurate models.

\vspace{0.135cm}

\begin{acknowledgments}
  We thank A.~P\'asztor for a careful reading of the manuscript, and
  both him and D.~Peszny\'ak for useful discussions.  This work was
  partially supported by NKFIH grants K-147396 and KKP-126769, and by
  the NKFIH excellence grant TKP2021-NKTA-64.
\end{acknowledgments}

\appendix

\begin{widetext}
  
\section{Gaussian integration}
\label{sec:app_gi}

To solve the model, we write the determinant $(\det\mathcal{M})^{N_f}$
in Eq.~\eqref{eq:Z} as a Grassmann integral,
\begin{equation}
  \label{eq:det}
  \begin{aligned}
    & \left[ \det
      \begin{pmatrix}
        m & iW + \mu \\ iW^\dag + \mu & m
      \end{pmatrix}
    \right]^{N_f} = \int d\Psi \int d\bar{\Psi} \int dX \int
    d\bar{X}\, e^{ m( \bar{X} X + \bar{\Psi} \Psi ) + \bar{X}(iW +
      \mu) \Psi + \bar{\Psi}(iW^\dag + \mu) X }\,,
  \end{aligned}
\end{equation}
where
\begin{equation}
  \label{eq:det_extra}
  \Psi=
  \begin{pmatrix}
    \psi^{(1)}\\ \ldots \\ \psi^{(N_f)}
  \end{pmatrix}\,, \qquad
  \bar{\Psi}=
  \begin{pmatrix}
    \bar{\psi}^{(1)}\\ \ldots \\ \bar{\psi}^{(N_f)}
  \end{pmatrix}\,, \qquad
    X=
  \begin{pmatrix}
    \chi^{(1)}\\ \ldots \\ \chi^{(N_f)}
  \end{pmatrix}\,, \qquad
    \bar{X}=
  \begin{pmatrix}
    \bar{\chi}^{(1)}\\ \ldots \\ \bar{\chi}^{(N_f)}
  \end{pmatrix}
\,,
\end{equation}
where each $\psi^{(f)},\bar{\psi}^{(f)},\chi^{(f)},\bar{\chi}^{(f)}$,
$f=1,\ldots, N_f$, is an $N$-component vector of Grassmann variables,
i.e., $\psi^{(f)}_i$, $i=1,\ldots, N$, and similarly for the others.
Summation over suppressed indices is understood. The partition
function Eq.~\eqref{eq:Z} reads then as
\begin{equation}
  \label{eq:Z_app}
  \begin{aligned}
    Z &= \int d\Psi \int d\bar{\Psi} \int dX \int d\bar{X}\, e^{ m(
      \bar{X} X + \bar{\Psi} \Psi ) + \mu( \bar{X} \Psi + \bar{\Psi}
      X) }
    \int d^2W  e^{-\act(W,W^\dag,\Psi,\bar{\Psi},X,\bar{X})}\,,\\
    \act & = N\left(\tr W W^\dag - \f{c}{N}|\tr W|^2\right) + i \tr W
    S + i \tr W^\dag \bar{S}\,,
  \end{aligned}
\end{equation}
where
\begin{equation}
  \label{eq:action_W}
  \begin{aligned}
    S_{ji} &= \tsum_f\psi^{(f)}_i \bar{\chi}^{(f)}_j\,, &&&
    \bar{S}_{ij} &= \tsum_f\chi^{(f)}_i \bar{\psi}^{(f)}_j\,.
  \end{aligned}
\end{equation}
Denoting $\mathbf{A}\cdot\mathbf{B} = \sum_{ij}A_{ij}B_{ij}$, we then
write
\begin{equation}
  \label{eq:action_W2}
  \begin{aligned}
    \act &= N \mathbf{W} \cdot (\Delta \mathbf{W}^{*}) + i \left(
      \mathbf{W} \cdot\mathbf{S} +
      \mathbf{W^{*}}\cdot\mathbf{\bar{S}}\right) = N \left(\mathbf{W}
      + \f{i}{N} \Delta^{-1}\mathbf{\bar{S}}\right) \cdot \left[\Delta
      \left( \mathbf{W}^{*} + \f{i}{N}\Delta^{-1} \mathbf{S}
      \right)\right] + \f{1}{N}\mathbf{\bar{S}}\cdot \Delta^{-1}
    \mathbf{S} \,,
  \end{aligned}
\end{equation}
where $(\Delta \mathbf{A})_{ij} = \sum_{kl}\Delta_{ij,kl} A_{kl}$ and
\begin{equation}
  \label{eq:delta_def}
  \Delta_{ij,kl} \equiv  \delta_{ik}\delta_{jl} -
  \f{c}{N}\delta_{ij}\delta_{kl}\,,\qquad
  \left(\Delta^{-1}\right)_{ij,kl} =  \delta_{ik}\delta_{jl} +
  \f{1}{N}\f{c}{1-c}\delta_{ij}\delta_{kl}\,.
\end{equation}
Integrating over $W$ we find
\begin{equation}
  \label{eq:action_W3}
  \begin{aligned}
    \int d^2W \, e^{-\act} & = \f{1}{\det
      \Delta}\left(\f{\pi}{N}\right)^N
    e^{-\f{1}{N}\mathbf{\bar{S}}\cdot \Delta^{-1} \mathbf{S} } \,,
  \end{aligned}
\end{equation}
where $\det \Delta = 1 -c$, and
\begin{equation}
  \label{eq:action_W6}
  \begin{aligned}
    - \mathbf{\bar{S}}\cdot \Delta^{-1} \mathbf{S} &= \trf C^{(\psi)}
    C^{(\chi)} -\f{1}{N}\f{c}{1-c} D^{\psi\chi}D^{\chi\psi}\,,
  \end{aligned}
\end{equation}
where
\begin{equation}
  \label{eq:action_W7}
  \begin{aligned}
    C^{(\psi)}_{fg} &= \tsum_{i}\bar{\psi}^{(f)}_i \psi^{(g)}_i\,,&&&
    C^{(\chi)}_{fg} &=  \tsum_{i}\bar{\chi}^{(f)}_i \chi^{(g)}_i \,,\\
    D^{\psi\chi}&=\tsum_{f,i}\bar{\psi}^{(f)}_i\chi^{(f)}_i\,,&&&
    D^{\chi\psi}&=\tsum_{f,i}\bar{\chi}^{(f)}_i \psi^{(f)}_i\,.
  \end{aligned}
\end{equation}
One proceeds to perform a Hubbard-Stratonovich transformation,
\begin{equation}
  \label{eq:action_W8}
  \begin{aligned}
    & \int d^2a \, e^{-N\trf\left( a^T a^*\right) + \trf\left( a^T
        C^{(\psi)}\right) + \trf\left( a^* C^{(\chi)}\right)} =
    \left(\f{\pi}{N}\right)^{N_f} e^{ \f{1}{N} \trf\left( C^{(\psi)}
        C^{(\chi)}\right)}\,,\\
    &\int d^2\omega \, e^{-N|\omega|^2 -
      \sqrt{\f{1}{N}\f{c}{1-c}}(\omega D^{\chi\psi} - \omega^*
      D^{\psi\chi})} = \f{\pi}{N} e^{- \f{1}{N^2} \f{c}{1-c}
      D^{\psi\chi}D^{\chi\psi}}\,,
  \end{aligned}
\end{equation}
where $a$ is a $N_f\times N_f$ complex matrix,
$d^2a = \prod_{ij}d \Re a_{ij} d\Im a_{ij}$, and
$d^2\omega = d\Re \omega d\Im\omega$, and explicitly
\begin{equation}
  \label{eq:action_W9}
  \begin{aligned}
    \trf\left( a^T C^{(\psi)}\right) + \trf \left(a^*
      C^{(\chi)}\right) &= \tsum_{i} \left[ \tsum_{f,g}
      \bar{\psi}^{(f)}_i (a)_{fg} \psi^{(g)}_i + \tsum_{f,g}
      \bar{\chi}^{(f)}_i (a^\dag)_{fg} \chi^{(g)}_i
    \right]\,,\\
    \omega D^{\chi\psi} - \omega^* D^{\psi\chi} &=
    \tsum_{i}\left(\omega \tsum_{f}\bar{\chi}^{(f)}_i \psi^{(f)}_i -
      \omega^* \tsum_{f}\bar{\psi}^{(f)}_i\chi^{(f)}_i\right)\,.
  \end{aligned}
\end{equation}
Collecting all the pieces that depend on Grassmann variables, we find
\begin{equation}
  \label{eq:action_Grass}
  \begin{aligned}
    & m(\bar{\Psi} \Psi + \bar{X} X) + \mu(\bar{X} \Psi + \bar{\Psi}X)
    + \trf\left( a^T C^{(\psi)}\right) + \trf\left( a^*
      C^{(\chi)}\right) - \sqrt{\f{1}{N}\f{c}{1-c}}( \omega
    D^{\chi\psi} - \omega^* D^{\psi\chi} )
    \\
    &= \tsum_i\bigg\{ \tsum_f m(\bar{\psi}^{(f)}_i \psi^{(f)}_i +
    \bar{\chi}^{(f)}_i \chi^{(f)}_i) + \mu(\bar{\chi}^{(f)}_i
    \psi^{(f)}_i + \bar{\psi}^{(f)}_i \chi^{(f)}_i) \\
    &\phantom{=\tsum_i}+ \left[ \tsum_{f,g} \bar{\psi}^{(f)}_i (a)_{fg}
      \psi^{(g)}_i + \tsum_{f,g} \bar{\chi}^{(f)}_i (a^\dag)_{fg}
      \chi^{(g)}_i\right] - \sqrt{\f{1}{N}\f{c}{1-c}} \left(\omega
      \tsum_{f}\bar{\chi}^{(f)}_i \psi^{(f)}_i - \omega^*
      \tsum_{f}\bar{\psi}^{(f)}_i\chi^{(f)}_i\right)\bigg\}\,.
\end{aligned}
\end{equation}
Performing now the Grassmann integration we conclude
\begin{equation}
  \label{eq:action_Grass2_app}
  Z =
  \f{\left({\pi}/{N}\right)^{N-N_f-1}}{\det \Delta}
  \int d^2a \int d^2\omega \, e^{-N\trf
    \left(a a^\dag\right) - N|\omega|^2}\left[  \det
    \begin{pmatrix}
      a  + m & \mu -   \sqrt{\f{1}{N}\f{c}{1-c}} \omega\\
      \mu + \sqrt{ \f{1}{N}\f{c}{1-c}} \omega^* & a^\dag + m
  \end{pmatrix}\right]^N\,,
\end{equation}
i.e., Eq.~\eqref{eq:action_Grass2}.

As mentioned in Sec.~\ref{sec:concl}, the trace deformation singles
out a direction in the integration manifold, i.e., the direction of
the matrix trace, that in the absence of the determinant becomes a
flat direction, on which the integrand does not depend, as $c$ tends
to 1. This is reminiscent of the flat directions corresponding to
gauge transformations when integrating over gauge fields in a gauge
theory, although the analogy is incomplete: (i) there is a remainder
that lifts the flat direction,
\begin{equation}
  \label{eq:trace2}
  \begin{aligned}
    \tr W W^\dag - \f{c}{N}|\tr W|^2 & = \tr \left(W-\f{1}{N}\tr
      W\right) \left(W-\f{1}{N}\tr W\right)^\dag + \f{1-c}{N}|\tr
    W|^2\,,
  \end{aligned}
\end{equation}
corresponding to what in a gauge theory is achieved by gauge fixing,
and (ii) the determinant also lifts the flat direction, which is not
the case in a gauge theory. In order to fully mimic the flat (gauge)
direction, we replace $W$ by $W-(b/N)\tr W$ in Eq.~\eqref{eq:det},
with $b,c\to 1$ as $N\to\infty$ to achieve the desired goal. The
resulting model can be solved through the same steps followed above,
leading to the same partition function
Eq.~\eqref{eq:action_Grass2_app} with the replacement
\begin{equation}
  \label{eq:app_flat_dir}
\f{c}{1-c} \to \f{(1-b)^2}{1-c} -1
\end{equation}
in the $\omega$-dependent contribution to the off-diagonal terms.  The
two most natural choices are $b=1$ and $b=c$. The first choice makes
the trace direction exactly flat for the determinant, in full analogy
with what happens in a gauge theory. With the second choice the flat
direction is still lifted, but in the same way as in the ``gauge''
action. In these two cases, however, the $\omega$-dependent
contribution is proportional to $\sqrt{-1/N}$ and $\sqrt{-c/N}$,
respectively, so it does not enter the saddle-point equations in the
large-$N$ limit, and one effectively ends up with the model of
Refs.~\cite{Jackson:1995nf,Stephanov:1996ki}. A nontrivial
contribution is obtained only if $(1-b)^2/(1-c) = \kappa^2 N$ (up to
subleading terms), which requires that $b,c$ approach 1 at different
(and related) rates. In this case one finds the same phase diagram
discussed in Sec.~\ref{sec:solut}.
\end{widetext}

\section{Solution of the model at $\mu=0$ and $m\neq 0$}
\label{sec:app_muzero_re}

At $\mu=0$, Eq.~\eqref{eq:sp_re2bis} for the $\Omega=0$ solutions
reduces to
\begin{equation}
  \label{eq:zeromu1}
  (A(A+m)-1)^2 (A+m)= 0\,.
\end{equation}
Since $Q=(A+m)^4$, the solution $A=-m$ gives $Q=0$ and must be
discarded. The other two solutions are
\begin{equation}
  \label{eq:zeromu2}
  \begin{aligned}
    A_\pm=\f{1}{2}\left( -m \pm \sqrt{m^2 +4}\right)\,,
  \end{aligned}
\end{equation}
and since $A$ and $m$ must have the same sign at the minimum, one has
that the plus (resp.\ minus) sign applies when $m>0$ (resp.\
$m<0$). The corresponding real part of the effective action reads as
\begin{equation}
  \label{eq:zeromu3}
  \begin{aligned}
    \mathcal{S}_\pm(m,\gamma) &\equiv \mathcal{S}(A_\pm,0,m,0,\gamma)
    \\
    &= \left(\tf{ -m \pm \sqrt{m^2 +4}}{2}\right)^2 -\ln \left(\tf{ m
        \pm \sqrt{m^2 +4}}{2}\right)^2\,.
  \end{aligned}
\end{equation}
When $\Omega\neq 0$, one has $Q = ((A+m)^2 + \Omega^2)^2\neq 0$, and
so one has to solve the set of equations
\begin{equation}
  \label{eq:zeromu4}
  \begin{aligned}
    & (A(1-\gamma)+m)\left((A+m)^2 +
      \Omega^2 \right)=0\,,\\
    & \left((A+m)^2 + \Omega^2 \right)\left((A+m)^2 + \Omega^2 -\gamma
    \right) = 0\,.
\end{aligned}
\end{equation}
There is no solution of this type for $\gamma=1$. For $\gamma\neq 1$
one has $\sqrt{Q}=\gamma$, and so
\begin{equation}
  \label{eq:zeromu5}
  \begin{aligned}
    A_0&= -\f{m}{1-\gamma}\,, &&& \Omega_0^2 & = \gamma\left(1 -
      m^2\f{\gamma}{(1-\gamma)^2}\right)\,.
  \end{aligned}
\end{equation}
Positivity of $\Omega_0^2$ requires
\begin{equation}
  \label{eq:zeromu6}
  (1-\gamma)^2> m^2 \gamma\,,
\end{equation}
while positivity of $m(A+m)$ requires $\gamma>1$, so we find that this
solution exists if
\begin{equation}
  \label{eq:zeromu7}
  \gamma >  \gamma_0(m) \equiv 1+\f{m^2}{2} + \sqrt{m^2 + \f{m^4}{4}} \,.
\end{equation}
The corresponding real part of the effective action reads as
\begin{equation}
  \label{eq:zeromu8}
  \mathcal{S}_0(m,\gamma) \equiv \mathcal{S}(A_0,\Omega_0,m,0,\gamma)
  = 1-\f{m^2}{\gamma-1} -\ln \gamma
  \,.
\end{equation}
To find the absolute minimum, set $2x = |m| + \sqrt{m^2 + 4}$.  It is
straightforward to show that $ x- \f{1}{x} = |m|$, and that
\begin{equation}
  \label{eq:zeromu12}
  y \equiv \f{x^2}{\gamma} =  \f{1}{2\gamma}\left(m^2 + 2 +
    |m|\sqrt{m^2+4}\right) 
  =\f{\gamma_0(m)}{\gamma}
  \,.
\end{equation}
We have
\begin{equation}
  \label{eq:zeromu10}
  \mathcal{S}_\pm -  \mathcal{S}_0 =
  \f{\gamma}{\gamma-1}y + \f{1}{\gamma-1}\f{1}{y}
  -  \f{\gamma+1}{\gamma-1} - \ln y\,,
\end{equation}
and
\begin{equation}
  \label{eq:zeromu11}
  \f{\de}{\de y} (\mathcal{S}_\pm -\mathcal{S}_0)  =
  \f{\gamma(y-1)}{(\gamma-1)y^2}\left(  y +\f{1}{\gamma}\right)\,. 
\end{equation}
When the solution $(A_0,\Omega_0)$ exists, i.e., when
Eq.~\eqref{eq:zeromu7} is satisfied, one has $0< y< 1$.  For
$\gamma>1$ and $0< y<1$ the right-hand side of Eq.~\eqref{eq:zeromu11}
is negative, and vanishes at $y=1$ where
$\mathcal{S}_\pm -\mathcal{S}_0$ is minimal (and vanishes), so that
$\mathcal{S}_\pm > \mathcal{S}_0$ whenever the solution with
$\Omega\neq 0$ exists.

\section{Solution of the model at $m=0$ -- real $\mu$}
\label{sec:app_mzero_re}

At $m=0$ and real chemical potential, for the $\Omega=0$ solutions one
has to solve
\begin{equation}
  \label{eq:sp_re2bismzero}
  \begin{aligned}
    A(A^2 - \mu^2)&= A\,.
  \end{aligned}
\end{equation}
There are three solutions, $(A_1,\Omega_1)$ and $(\pm A_2,\Omega_2)$,
with $\Omega_{1,2}=0$ and
\begin{equation}
  \label{eq:sp_re2bismzero_sol}
  \begin{aligned}
    A_1=0\,, \qquad A_2 = \sqrt{1+\mu^2}\,,
  \end{aligned}
\end{equation}
that exist $\forall \mu,\gamma$. The corresponding effective action
$\mathcal{S}(A,\Omega,0,\mu,\gamma)$ reads as
\begin{equation}
  \label{eq:sp_re2bismzero_sol_effact}
  \begin{aligned}
    {\cal S}^{(1)}(\mu,\gamma) &= \mathcal{S}(0,0,0,\mu,\gamma) =    -\ln\mu^2 \,,\\
    {\cal S}^{(2)}(\mu,\gamma) &= \mathcal{S}(\pm A_2,0,0,\mu,\gamma)
    = 1 + \mu^2 \\ &= {\cal S}^{(1)}(\mu,\gamma) + f_1(\mu^2)\,,
  \end{aligned}
\end{equation}
where $f_1(x) = 1 + x+ \ln x$. For the $\Omega\neq 0$ solutions one
has to solve instead
\begin{equation}
  \label{eq:sp_re2termzero}
  \begin{aligned}
    A\gamma\left(A^2 + \Omega^2 + \mu^2\right) &= A\left(A^2 +
      \Omega^2 - \mu^2\right)\,,\\
    \left(A^2 + \Omega^2 - \mu^2\right)^2 + 4\mu^2\Omega^2 &
    =\gamma\left(A^2 + \Omega^2 + \mu^2\right) \,,
  \end{aligned}
\end{equation}
so there are $A=0$ and $A\neq 0$ solutions.  If $A=0$ the second
equation becomes
\begin{equation}
  \label{eq:sp_re2termzero1}
  \begin{aligned}
    \left(\Omega^2 + \mu^2\right)\left(\Omega^2 +
      \mu^2-\gamma\right)&=0\,,
  \end{aligned}
\end{equation}
and one has the two solutions $(A_3,\pm \Omega_3)$ with $A_3=0$ and
\begin{equation}
  \label{eq:sp_re2termzero1bis}
  \begin{aligned}
    \Omega_3 = \sqrt{\gamma-\mu^2}\,,
  \end{aligned}
\end{equation}
which exist if $\gamma>\mu^2$. The corresponding effective action is
\begin{equation}
  \label{eq:sp_re2termzero1bis_eff}
  \begin{aligned}
    {\cal S}^{(3)}(\mu,\gamma) &= \mathcal{S}(0,\pm
    \Omega_3,0,\mu,\gamma) = 1- \f{\mu^2}{\gamma} +\ln \f{1}{\gamma}
    \\ &    = {\cal S}^{(1)}(\mu,\gamma)  + f_2\left(\tf{\mu^2}{\gamma}\right)\\
    &= {\cal S}^{(2)}(\mu,\gamma) +
    f_3\left(\mu^2,\tf{1}{\gamma}\right) \,,
  \end{aligned}
\end{equation}
where
\begin{equation}
  \label{eq:flist}
  \begin{aligned}
    f_2(x)&=1- x +\ln x\,, \\
    f_3(x,y)&= - x\left(y+1\right) +\ln y\,.
  \end{aligned}
\end{equation}
Finally, if $A\neq 0$ the first equation in
Eq.~\eqref{eq:sp_re2termzero} implies
\begin{equation}
  \label{eq:sp_re2termzero2}
  \begin{aligned}
    A^2 + \Omega^2   &= \f{1+\gamma}{1-\gamma}\mu^2\,,
  \end{aligned}
\end{equation}
which has a solution only for $\gamma<1$. Plugging this into the
second equation in Eq.~\eqref{eq:sp_re2termzero}, one finds the four
solutions $(\pm A_4,\pm \Omega_4)$, with
\begin{equation}
  \label{eq:sp_re2termzero3}
  \begin{aligned}
    \Omega_4^2 & =\f{\gamma}{2(1-\gamma)} -
    \f{\gamma^2\mu^2}{(1-\gamma)^2} \,,
  \end{aligned}
\end{equation}
and
\begin{equation}
  \label{eq:sp_re2termzero4}
  \begin{aligned}
    A_4^2 &= -\f{\gamma}{2(1-\gamma)} + \f{\mu^2}{(1-\gamma)^2}\,.
\end{aligned}
\end{equation}
These solutions exist only if $\gamma<1$ and
\begin{equation}
  \label{eq:sp_re2termzero5}
  \begin{aligned}
    \f{\gamma(1-\gamma)}{2}< \mu^2 < \f{1-\gamma}{2\gamma}\,.
  \end{aligned}
\end{equation}
The corresponding effective action reads as
\begin{equation}
  \label{eq:sp_re2termzero5_eff}
  \begin{aligned}
    {\cal S}^{(4)}(\mu,\gamma) &= \mathcal{S}(\pm A_4,\pm
    \Omega_4,0,\mu,\gamma)\\ & = \f{1}{2}\left( 1 +\f{2\mu^2
      }{1-\gamma} - \ln\f{2\mu^2\gamma}{1-\gamma}\right) \\ &= {\cal
      S}^{(2)}(\mu,\gamma) -
    \f{1}{2}f_2\left(\tf{2\mu^2\gamma}{1-\gamma}\right) \\
    &= {\cal S}^{(3)}(\mu,\gamma) -
    \f{1}{2}f_2\left(\tf{2\mu^2}{\gamma(1-\gamma)}\right)\,.
  \end{aligned}
\end{equation}
One has $f_1(x)\le 0$ for $0\le x \le x_*\simeq 0.2785$, with
$f_1(x_*)=0$, and $f_2(x)\le 0$ $\forall x$, with $f_2(1)=0$; these
are the only real positive zeros of these functions. Then certainly
${\cal S}^{(4)}> {\cal S}^{(2)}$ or ${\cal S}^{(4)}> {\cal S}^{(3)}$
when solution 4 exists, so it can always be ignored.  Next,
${\cal S}^{(2)}\le {\cal S}^{(1)}$ for $\mu^2\le x_*$, with equality
holding only if $\mu^2=x_*$; and ${\cal S}^{(3)}\le {\cal S}^{(1)}$
for $\mu^2\le \gamma$, with equality holding only if $\mu^2=\gamma$,
so that ${\cal S}^{(3)}< {\cal S}^{(1)}$ whenever solution 3
exists. Finally, ${\cal S}^{(3)}\le {\cal S}^{(2)}$ when
\begin{equation}
  \label{eq:saddle_point12}
  - \left(\f{1}{\gamma}+1\right)\mu^2 +\ln \f{1}{\gamma} \le 0\,.  
\end{equation}
If $\gamma\ge 1$ this inequality is certainly satisfied.  If
$0<\gamma<1$, together with the request that solution 3 exists, one
has
\begin{equation}
  \label{eq:saddle_point13}
  b(\gamma)\equiv    \f{\gamma\ln\f{1}{\gamma}}{1+\gamma} \le \mu^2 \le \gamma\,.
\end{equation}
A window is available when $\ln\f{1}{\gamma}\le 1+\gamma$, which is
the case for $\gamma\ge \gamma_*$, with $\gamma_*$ satisfying
\begin{equation}
  \label{eq:saddle_point14}
  1 + \gamma_* + \ln \gamma_* = f_1(\gamma_*)= 0\,,
\end{equation}
i.e., $\gamma_* = x_*$. We conclude that
\begin{equation}
  \label{eq:free_energy}
  \mathcal{F}  =\left\{
    \begin{aligned}
      & 1+ \mu^2\,, &&& &\mu^2\le\min(x_*,b(\gamma))\,,\\
      & 1- \tf{\mu^2}{\gamma} -\ln \gamma\,, &&& &b(\gamma)\le \mu^2
      \le \gamma\,, \\
      & -\ln\mu^2\,, &&& &\mu^2 \ge \max(x_*,\gamma)\,.
    \end{aligned}\right.
\end{equation}
For the first derivatives we find
\begin{equation}
  \label{eq:free_energy2}
  \begin{aligned}
    \mathcal{F}_\mu &= \left\{
    \begin{aligned}
      & 2\mu\,,\\
      & -\tf{2\mu}{\gamma} \,,
      \\
      & -\tf{2}{\mu}\,,
    \end{aligned}\right.
  &
  \begin{aligned}
    &\mu^2<\min(x_*,b(\gamma))\,,\\
    &b(\gamma)< \mu^2
    < \gamma\,, \\
    &\mu^2 > \max(x_*,\gamma)\,,
    \end{aligned}
    \\
    \mathcal{F}_\gamma &= \left\{
    \begin{aligned}
      & 0\,,\\
      & \tf{\mu^2-\gamma}{\gamma^2}
      \,, \\
      & 0\,,
    \end{aligned}\right.
  &
  \begin{aligned}
    &\mu^2<\min(x_*,b(\gamma))\,,\\
    &b(\gamma)< \mu^2
    < \gamma\,, \\
    &\mu^2 > \max(x_*,\gamma)\,,
  \end{aligned}
\end{aligned}
\end{equation}
while for the second derivatives we find
\begin{equation}
  \label{eq:free_energy3}
  \begin{aligned}
    \mathcal{F}_{\mu\mu}&= \left\{
    \begin{aligned}
      & 2\,,\\
      & -\tf{2}{\gamma} \,,\\
      & \tf{2}{\mu^2}\,,
    \end{aligned}\right.
  &
  \begin{aligned}
    &\mu^2<\min(x_*,b(\gamma))\,,\\
    &b(\gamma)< \mu^2
    < \gamma\,, \\
    &\mu^2 > \max(x_*,\gamma)\,,
  \end{aligned}
  \\
  \mathcal{F}_{\gamma\gamma} &= \left\{
    \begin{aligned}
      & 0\,,\\
      & \tf{\gamma-2\mu^2}{\gamma^3}
      \,, \\
      & 0\,,
    \end{aligned}\right.
  &
  \begin{aligned}
    &\mu^2<\min(x_*,b(\gamma))\,,\\
    &b(\gamma)< \mu^2
    < \gamma\,, \\
    &\mu^2 > \max(x_*,\gamma)\,,
    \end{aligned}
    \\
    \mathcal{F}_{\mu\gamma} &= \left\{
    \begin{aligned}
      & 0\,, \\
      & \tf{2\mu}{\gamma^2}
      \,, \\
      & 0\,,
    \end{aligned}\right.
  &
  \begin{aligned}
    &\mu^2<\min(x_*,b(\gamma))\,,\\
    &b(\gamma)< \mu^2
    < \gamma\,, \\
    &\mu^2 > \max(x_*,\gamma)\,.
    \end{aligned}
  \end{aligned}
\end{equation}
One can define also the mass derivative at $m=0$ as a limit,
$\mathcal{F}_m^{(\pm)}\equiv \lim_{m\to 0^\pm}\mathcal{F}_m(m) =
-2\lim_{m\to 0^\pm}A_0(m) $, which reads
\begin{equation}
  \label{eq:app_massder}
  \begin{aligned}
    \mathcal{F}_m^{(\pm)} &= \left\{
    \begin{aligned}
      & \mp 2\sqrt{1+\mu^2}\,, \\ 
      & 0 \,, \\ 
      & 0\,,
    \end{aligned}\right.
  &
  \begin{aligned}
    &\mu^2<\min(x_*,b(\gamma))\,,\\
    &b(\gamma)< \mu^2
    < \gamma\,, \\
    &\mu^2 > \max(x_*,\gamma)\,.
    \end{aligned}
  \end{aligned}
\end{equation}
At $\mu^2=x_*$ and $\gamma<x_*$, below the triple point
$\mu^2=\gamma=x_*$ (transition line $L_1$), the transition is first
order with $\mathcal{F}_\mu$ and $\mathcal{F}_m^{(\pm)}$ discontinuous
(and $\mathcal{F}_\gamma$ continuous as it vanishes on both sides).
For $\mu^2=b(\gamma)<x_*$ (transition line $L_2$), one has a
discontinuous $\mathcal{F}_m^{(\pm)}$, and moreover
\begin{equation}
  \label{eq:delta1}
  \begin{aligned}
    \Delta \mathcal{F}_\mu|_{\mu^2=b(\gamma)}& \equiv
    \mathcal{F}_\mu|_{\mu^2\to b(\gamma)^+} -
    \mathcal{F}_\mu|_{\mu^2\to b(\gamma)^-} \\ & = \f{-2\mu}{\gamma}
    -2\mu \\
    &=-\mathrm{sgn}(\mu)\sqrt{\left(1+\f{1}{\gamma}\right)\ln\f{1}{\gamma}}\,,
  \end{aligned}
\end{equation}
and
\begin{equation}
  \label{eq:delta3}
  \begin{aligned}
    \Delta \mathcal{F}_\gamma|_{\mu^2=b(\gamma)} =
    \f{\f{\gamma}{1+\gamma}\ln\f{1}{\gamma}-\gamma}{\gamma^2} =
    -\f{f_1(\gamma)}{\gamma(1+\gamma)}\,.
\end{aligned}
\end{equation}
One has that $\mathcal{F}_\mu$ is discontinuous along the whole line
$L_2$ except at $(\mu=0,\gamma=1)$, where, however,
\begin{equation}
  \label{eq:delta2}
  \Delta   \mathcal{F}_{\mu\mu}
  |_{\mu^2=b(\gamma)\to 0}
  = -4\,.
\end{equation}
On the other hand, $\mathcal{F}_\gamma$ is discontinuous along the
whole transition line $L_2$ except at the triple point
$\mu^2=\gamma=x_*$ where $f_1(x_*)=0$, but where
\begin{equation}
  \label{eq:delta2bis}
  \Delta  \mathcal{F}_{\gamma\gamma}
  |_{\mu^2=b(\gamma)\to x_*}
  = -\f{1}{x_*^2}\,.
\end{equation}
The transition is then first order along the whole line $L_2$, with
peculiar behavior at its two extremes. Finally, at the transition
across the line $\mu^2=\gamma$ for $\gamma>x_*$ (transition line
$L_3$), $\mathcal{F}_\mu$ and $\mathcal{F}_\gamma$ (as well as
$\mathcal{F}_m^{(\pm)}$) are continuous, while $\mathcal{F}_{\mu\mu}$,
$\mathcal{F}_{\gamma\gamma}$, and $\mathcal{F}_{\mu\gamma}$ are not
(this behavior persists also as one approaches the triple point), so
the transition is second order on $L_3$.

For the imaginary part of the action one finds
\begin{equation}
  \label{eq:eff_act_phase_saddle2}
  \varphi= \left\{
  \begin{aligned}
    &0\,, &&& &\mu^2\le\min(x_*,b(\gamma))\,,\\
    & \mathrm{sgn}(\Omega_0)\, 2\arcsin\tf{ \mu}{\sqrt{\gamma}}\,, &&&
    &b(\gamma)\le \mu^2 \le \gamma\,,
    \\
    & \pm\pi\,, &&& &\mu^2 \ge \max(x_*,\gamma)\,,
  \end{aligned}
\right.
\end{equation}
with $\arcsin(x)\in \left[-\tf{\pi}{2},\tf{\pi}{2}\right]$. The sign
of $\Omega_0$, and the sign of $\varphi=\pm\pi$ for
$\mu^2 \ge \max(x_*,\gamma)$, are opposite to the sign of $\mu_I$ as
this approaches zero. The second line follows from the fact that
$\mathrm{sgn}(\sin\Phi) = \mathrm{sgn}(\mu\Omega_0)$ and
\begin{equation}
  \label{eq:eff_act_phase_saddle2_bis}
  \begin{aligned}
    \cos\varphi &= 1- 2\left(\sin\tf{\varphi}{2}\right)^2 = 1-\f{ 2
      \mu^2}{\gamma}\,, \quad\text{if }b(\gamma)\le \mu^2 \le
    \gamma\,,
  \end{aligned}
\end{equation}
see Eq.~\eqref{eq:sp7}. The phase $\varphi$ changes discontinuously at
the transition along the line $L_2$ ($\mu^2=b(\gamma)$, $\mu^2<x_*$),
where
\begin{equation}
  \label{eq:deltaphi}
  \Delta( \cos \varphi)\big|_{\mu^2=b(\gamma)}
  =- \tf{2}{1+\gamma}\ln\tf{1}{\gamma}\,. 
\end{equation}
The phase is discontinuous also across the line $L_1$ ($\mu^2=x_*$,
$\gamma< x_*$), where it jumps from 0 to $\pm\pi$.  The phase changes
continuously across the transition line $L_3$ ($\mu^2=\gamma$,
$\mu^2>x_*$), where, however,
\begin{equation}
  \label{eq:deltaphi2}
  \f{\de}{\de \mu^2}\cos
  \varphi\Big|_{\mu^2\to \gamma^+} -
  \f{\de}{\de \mu^2}\cos
  \varphi\Big|_{\mu^2\to \gamma^-} 
  = \f{2}{\mu^2}\,.
\end{equation}

\section{Solution of the model at $m=0$ -- imaginary $\mu$}
\label{sec:app_mzero_im}

For purely imaginary chemical potential at vanishing mass, the
saddle-point equations simplify to
\begin{equation}
  \label{eq:sp_im2_m0}
  \begin{aligned}
    0 &= A(Q_I - 1)\,,\\
    0&= \Omega Q_I - \gamma(\Omega-\mu_I)\,,
  \end{aligned}
\end{equation}
with $Q_I= A^2 + (\Omega-\mu_I)^2$. There are two types of solutions:
$A=0$, and $A\neq 0$. If $A=0$ one has
\begin{equation}
  \label{eq:sp_im2_m0_1}
  (\Omega-\mu_I)\left(  \Omega(\Omega-\mu_I) - \gamma\right) =0\,,
\end{equation}
which since $\Omega=\mu_I$ implies $Q=0$, has as only acceptable
solutions
\begin{equation}
  \label{eq:sp_im2_m0_2}
  \Omega_\pm = \f{\mu_I \pm \sqrt{\mu_I^2 + 4\gamma}}{2}\,, 
\end{equation}
with $\Omega_\pm \gtrless 0$, and corresponding action
\begin{equation}
  \label{eq:sp_im2_m0_2bis}
  \begin{aligned}
    \mathcal{S}_I^{(\mp)}(\mu_I,\gamma) &\equiv
    \mathcal{S}_I(0,\Omega_\mp,0,\mu_I,\gamma) \\ &=
    \f{1}{\gamma}\left(\tf{\mu_I \mp \sqrt{\mu_I^2 +
          4\gamma}}{2}\right)^2 - \ln\left(\tf{-\mu_I \mp
        \sqrt{\mu_I^2 + 4\gamma}}{2}\right)^2\,.
  \end{aligned}
\end{equation}
We know from general arguments [see after Eq.~\eqref{eq:9} in
Sec.~\ref{sec:immu}] that for these solutions
$\mathcal{S}_I^{(-)}< \mathcal{S}_I^{(+)}$ (resp.\
$\mathcal{S}_I^{(-)}> \mathcal{S}_I^{(+)}$) if $\mu_I> 0$ (resp.\
$\mu_I< 0$). If $A\neq 0$ instead one must have $Q_I=1$, so from the
second equation in Eq.~\eqref{eq:sp_im2_m0} one finds
\begin{equation}
  \label{eq:sp_im2_m0_3}
  \Omega_0  = -\f{\gamma\mu_I }{1-\gamma}\,,
\end{equation}
which plugged back into $Q_I$ gives
\begin{equation}
  \label{eq:sp_im2_m0_4}
  A_0^2 = 1-\f{\mu_I^2}{(1-\gamma)^2}\,.
\end{equation}
This solution exists only if $\mu_I^2 < (1-\gamma)^2$.  The
corresponding action is
\begin{equation}
  \label{eq:sp_im2_m0_4_bis}
  \begin{aligned}
    \mathcal{S}_I^{(0)}(\mu_I,\gamma) &= \mathcal{S}_I(\pm
    A_0,\Omega_0,0,\mu_I,\gamma) = 1 - \f{\mu_I^2}{1-\gamma} \,.
  \end{aligned}
\end{equation}
To find the absolute minimum we need only compare
$\mathcal{S}_I^{(\mp)}$ with $\mathcal{S}_I^{(0)}$; since
$\mathcal{S}_I^{(+)}(-\mu_I,\gamma)=\mathcal{S}_I^{(-)}(\mu_I,\gamma)$,
it suffices to choose $\mu_I>0$ and compare $\mathcal{S}_I^{(-)}$ with
$\mathcal{S}_I^{(0)}$. Set
\begin{equation}
  \label{eq:im_mu18_0}
  \begin{aligned}
    x&=\f{\sqrt{\mu_I^2+4\gamma}+\mu_I}{2}\ge 0\,.
  \end{aligned}
\end{equation}
Notice first that $\mathcal{S}_I^{(-)} = \f{\gamma}{x^2} - \ln x^2$,
and
\begin{equation}
  \label{eq:im_mu18_0_bis}
  \f{\de \mathcal{S}_I^{(-)}}{\de x} = -2\left(\f{\gamma}{x^3}
    +\f{1}{x}\right) 
  < 0
  \quad \text{if } x> 0\,;
\end{equation}
at $\mu_I=0$, $\mathcal{S}_I^{(-)}=1-\ln \gamma$, so for $\mu_I\ge 0$
and $\gamma\ge 1$, one has $\mathcal{S}_I^{(-)}\le 1$ while
$\mathcal{S}_I^{(0)}\ge 1$, so
$\mathcal{S}_I^{(0)}\ge \mathcal{S}_I^{(-)}$. Next, since
$(2x- \mu_I)^2 = \mu_I^2 + 4\gamma$, we have
$ \mu_I = x - \f{\gamma}{x}$, from which follows
\begin{equation}
  \label{eq:im_mu18_quinquies}
  \mathcal{S}_I^{(0)}-\mathcal{S}_I^{(-)} 
  =  - \f{\gamma+1}{\gamma-1} + \f{1}{\gamma-1}z
  +\f{1}{\gamma-1}
  \f{\gamma}{z} + \ln z\,,
\end{equation}
where $z=x^2$, and so
\begin{equation}
  \label{eq:im_mu18_sexties}
  \f{\de}{\de z}(\mathcal{S}_I^{(0)}-\mathcal{S}_I^{(-)}) =
  \f{(1-z)(z+\gamma)}{(1-\gamma)z^2}\,.
\end{equation}
For $0<\gamma< 1$ and $z>0$, this shows that
$\mathcal{S}_I^{(0)}-\mathcal{S}_I^{(-)}$ has its maximum at $z=1$;
since $\mathcal{S}_I^{(0)}=\mathcal{S}_I^{(-)}$ there,
$\mathcal{S}_I^{(0)}$ is the minimum whenever the corresponding
solution exists. Summarizing, $\mathcal{F}=\mathcal{S}_I^{(\mp)}$ if
$\pm\mu_I \ge \max(0,1-\gamma)$, and $\mathcal{F}=\mathcal{S}_I^{(0)}$
if $|\mu_I| \le 1-\gamma$. There is then a line of transitions at
$\mu_I=1-\gamma$, one at $-\mu_I=1-\gamma$, and one at $\mu_I=0$ for
$\gamma>1$.  Using the following expressions for the derivatives,
\begin{equation}
  \label{eq:im_mu_der1}
  \begin{aligned}
    \f{\de \mathcal{S}_I^{(\mp)}}{\de \mu_I} &=
    \f{1}{\gamma}\left(\mu_I \mp \sqrt{\mu_I^2 + 4\gamma}\right)\,,\\
    \f{\de^2 \mathcal{S}_I^{(\mp)}}{\de \mu_I^2} &=
    \f{1}{\gamma}\left(1 \mp \f{\mu_I}{\sqrt{\mu_I^2 + 4\gamma}}\right)\,,\\
    \f{\de \mathcal{S}_I^{(\mp)} }{\de \gamma} &=
    -\f{\mu_I}{2\gamma^2}\left(\mu_I \mp \sqrt{\mu_I^2+4\gamma}\right) - \f{1}{\gamma}\,,\\
    \f{\de^2 \mathcal{S}_I^{(\mp)}}{\de \gamma^2} &=
    \f{1}{\gamma^3}\left( \gamma+ \mu_I^2 \mp \f{\mu_I(\mu_I^2 + 3
        \gamma)}{ \sqrt{\mu_I^2 + 4 \gamma}} \right)\,,
  \end{aligned}
\end{equation}
and
\begin{equation}
  \label{eq:im_mu_der1_bis}
  \begin{aligned}
    \f{\de \mathcal{S}_I^{(0)}}{\de \mu_I} &=-\f{2\mu_I}{1-\gamma}\,,
    &&& \f{\de^2 \mathcal{S}_I^{(0)}}{\de \mu_I^2}
    &=-\f{2}{1-\gamma}\,,\\
    \f{\de \mathcal{S}_I^{(0)}}{\de \gamma}
    &=-\f{\mu_I^2}{(1-\gamma)^2}\,,&&& \f{\de^2
      \mathcal{S}_I^{(0)}}{\de \gamma^2} &=
    -\f{2\mu_I^2}{(1-\gamma)^3}\,,
  \end{aligned}
\end{equation}
we find that $\mathcal{F}_{\mu_I}$ is continuous on the first two
transition lines but discontinuous on the third one, since
\begin{equation}
  \label{eq:im_mu_der2}
  \begin{aligned}
    \f{\de }{\de \mu_I}
    \left(\mathcal{S}_I^{(-)}-\mathcal{S}_I^{(+)}\right)|_{\mu_I =0}
    &= -\f{4}{\sqrt{\gamma}}\,,
  \end{aligned}
\end{equation}
while $\mathcal{F}_\gamma$ is continuous on all three lines. On the
other hand, $\mathcal{F}_{\mu_I\mu_I}$ and
$\mathcal{F}_{\gamma\gamma}$ are discontinuous on the first two
transition lines.

\mbox{}

\bibliographystyle{apsrev4-2}
\bibliography{references_trdef}

\begin{thebibliography}{46}%
\makeatletter
\providecommand \@ifxundefined [1]{%
 \@ifx{#1\undefined}
}%
\providecommand \@ifnum [1]{%
 \ifnum #1\expandafter \@firstoftwo
 \else \expandafter \@secondoftwo
 \fi
}%
\providecommand \@ifx [1]{%
 \ifx #1\expandafter \@firstoftwo
 \else \expandafter \@secondoftwo
 \fi
}%
\providecommand \natexlab [1]{#1}%
\providecommand \enquote  [1]{``#1''}%
\providecommand \bibnamefont  [1]{#1}%
\providecommand \bibfnamefont [1]{#1}%
\providecommand \citenamefont [1]{#1}%
\providecommand \href@noop [0]{\@secondoftwo}%
\providecommand \href [0]{\begingroup \@sanitize@url \@href}%
\providecommand \@href[1]{\@@startlink{#1}\@@href}%
\providecommand \@@href[1]{\endgroup#1\@@endlink}%
\providecommand \@sanitize@url [0]{\catcode `\\12\catcode `\$12\catcode
  `\&12\catcode `\#12\catcode `\^12\catcode `\_12\catcode `\%12\relax}%
\providecommand \@@startlink[1]{}%
\providecommand \@@endlink[0]{}%
\providecommand \url  [0]{\begingroup\@sanitize@url \@url }%
\providecommand \@url [1]{\endgroup\@href {#1}{\urlprefix }}%
\providecommand \urlprefix  [0]{URL }%
\providecommand \Eprint [0]{\href }%
\providecommand \doibase [0]{https://doi.org/}%
\providecommand \selectlanguage [0]{\@gobble}%
\providecommand \bibinfo  [0]{\@secondoftwo}%
\providecommand \bibfield  [0]{\@secondoftwo}%
\providecommand \translation [1]{[#1]}%
\providecommand \BibitemOpen [0]{}%
\providecommand \bibitemStop [0]{}%
\providecommand \bibitemNoStop [0]{.\EOS\space}%
\providecommand \EOS [0]{\spacefactor3000\relax}%
\providecommand \BibitemShut  [1]{\csname bibitem#1\endcsname}%
\let\auto@bib@innerbib\@empty
\bibitem [{\citenamefont {Guenther}(2022)}]{Guenther:2022wcr}%
  \BibitemOpen
  \bibfield  {author} {\bibinfo {author} {\bibfnamefont {J.~N.}\ \bibnamefont
  {Guenther}},\ }\href {https://doi.org/10.22323/1.396.0013} {\bibfield
  {journal} {\bibinfo  {journal} {PoS}\ }\textbf {\bibinfo {volume}
  {LATTICE2021}},\ \bibinfo {pages} {013} (\bibinfo {year} {2022})},\ \Eprint
  {https://arxiv.org/abs/2201.02072} {arXiv:2201.02072 [hep-lat]} \BibitemShut
  {NoStop}%
\bibitem [{\citenamefont {Aarts}\ \emph {et~al.}(2023)\citenamefont {Aarts}
  \emph {et~al.}}]{Aarts:2023vsf}%
  \BibitemOpen
  \bibfield  {author} {\bibinfo {author} {\bibfnamefont {G.}~\bibnamefont
  {Aarts}} \emph {et~al.},\ }\href {https://doi.org/10.1016/j.ppnp.2023.104070}
  {\bibfield  {journal} {\bibinfo  {journal} {Prog. Part. Nucl. Phys.}\
  }\textbf {\bibinfo {volume} {133}},\ \bibinfo {pages} {104070} (\bibinfo
  {year} {2023})},\ \Eprint {https://arxiv.org/abs/2301.04382}
  {arXiv:2301.04382 [hep-lat]} \BibitemShut {NoStop}%
\bibitem [{\citenamefont {P{\'a}sztor}(2024)}]{Pasztor:2024dpv}%
  \BibitemOpen
  \bibfield  {author} {\bibinfo {author} {\bibfnamefont {A.}~\bibnamefont
  {P{\'a}sztor}},\ }\href {https://doi.org/10.22323/1.453.0108} {\bibfield
  {journal} {\bibinfo  {journal} {PoS}\ }\textbf {\bibinfo {volume}
  {LATTICE2023}},\ \bibinfo {pages} {108} (\bibinfo {year} {2024})}\BibitemShut
  {NoStop}%
\bibitem [{\citenamefont {Torres-Rincon}\ and\ \citenamefont
  {Aichelin}(2017)}]{Torres-Rincon:2017zbr}%
  \BibitemOpen
  \bibfield  {author} {\bibinfo {author} {\bibfnamefont {J.~M.}\ \bibnamefont
  {Torres-Rincon}}\ and\ \bibinfo {author} {\bibfnamefont {J.}~\bibnamefont
  {Aichelin}},\ }\href {https://doi.org/10.1103/PhysRevC.96.045205} {\bibfield
  {journal} {\bibinfo  {journal} {Phys. Rev. C}\ }\textbf {\bibinfo {volume}
  {96}},\ \bibinfo {pages} {045205} (\bibinfo {year} {2017})},\ \Eprint
  {https://arxiv.org/abs/1704.07858} {arXiv:1704.07858 [nucl-th]} \BibitemShut
  {NoStop}%
\bibitem [{\citenamefont {Braun}\ \emph {et~al.}(2017)\citenamefont {Braun},
  \citenamefont {Leonhardt},\ and\ \citenamefont {Pospiech}}]{Braun:2017srn}%
  \BibitemOpen
  \bibfield  {author} {\bibinfo {author} {\bibfnamefont {J.}~\bibnamefont
  {Braun}}, \bibinfo {author} {\bibfnamefont {M.}~\bibnamefont {Leonhardt}},\
  and\ \bibinfo {author} {\bibfnamefont {M.}~\bibnamefont {Pospiech}},\ }\href
  {https://doi.org/10.1103/PhysRevD.96.076003} {\bibfield  {journal} {\bibinfo
  {journal} {Phys. Rev. D}\ }\textbf {\bibinfo {volume} {96}},\ \bibinfo
  {pages} {076003} (\bibinfo {year} {2017})},\ \Eprint
  {https://arxiv.org/abs/1705.00074} {arXiv:1705.00074 [hep-ph]} \BibitemShut
  {NoStop}%
\bibitem [{\citenamefont {Kov\'acs}\ \emph {et~al.}(2016)\citenamefont
  {Kov\'acs}, \citenamefont {Sz\'ep},\ and\ \citenamefont
  {Wolf}}]{Kovacs:2016juc}%
  \BibitemOpen
  \bibfield  {author} {\bibinfo {author} {\bibfnamefont {P.}~\bibnamefont
  {Kov\'acs}}, \bibinfo {author} {\bibfnamefont {{\relax Zs}.}~\bibnamefont
  {Sz\'ep}},\ and\ \bibinfo {author} {\bibfnamefont {{\relax Gy}.}~\bibnamefont
  {Wolf}},\ }\href {https://doi.org/10.1103/PhysRevD.93.114014} {\bibfield
  {journal} {\bibinfo  {journal} {Phys. Rev. D}\ }\textbf {\bibinfo {volume}
  {93}},\ \bibinfo {pages} {114014} (\bibinfo {year} {2016})},\ \Eprint
  {https://arxiv.org/abs/1601.05291} {arXiv:1601.05291 [hep-ph]} \BibitemShut
  {NoStop}%
\bibitem [{\citenamefont {Alford}\ \emph {et~al.}(1998)\citenamefont {Alford},
  \citenamefont {Rajagopal},\ and\ \citenamefont {Wilczek}}]{Alford:1997zt}%
  \BibitemOpen
  \bibfield  {author} {\bibinfo {author} {\bibfnamefont {M.~G.}\ \bibnamefont
  {Alford}}, \bibinfo {author} {\bibfnamefont {K.}~\bibnamefont {Rajagopal}},\
  and\ \bibinfo {author} {\bibfnamefont {F.}~\bibnamefont {Wilczek}},\ }\href
  {https://doi.org/10.1016/S0370-2693(98)00051-3} {\bibfield  {journal}
  {\bibinfo  {journal} {Phys. Lett. B}\ }\textbf {\bibinfo {volume} {422}},\
  \bibinfo {pages} {247} (\bibinfo {year} {1998})},\ \Eprint
  {https://arxiv.org/abs/hep-ph/9711395} {arXiv:hep-ph/9711395} \BibitemShut
  {NoStop}%
\bibitem [{\citenamefont {Casalbuoni}(2006)}]{Casalbuoni:2006rs}%
  \BibitemOpen
  \bibfield  {author} {\bibinfo {author} {\bibfnamefont {R.}~\bibnamefont
  {Casalbuoni}},\ }\href {https://doi.org/10.22323/1.029.0001} {\bibfield
  {journal} {\bibinfo  {journal} {PoS}\ }\textbf {\bibinfo {volume}
  {CPOD2006}},\ \bibinfo {pages} {001} (\bibinfo {year} {2006})},\ \Eprint
  {https://arxiv.org/abs/hep-ph/0610179} {arXiv:hep-ph/0610179} \BibitemShut
  {NoStop}%
\bibitem [{\citenamefont {Verbaarschot}\ and\ \citenamefont
  {Wettig}(2000)}]{Verbaarschot:2000dy}%
  \BibitemOpen
  \bibfield  {author} {\bibinfo {author} {\bibfnamefont {J.~J.~M.}\
  \bibnamefont {Verbaarschot}}\ and\ \bibinfo {author} {\bibfnamefont
  {T.}~\bibnamefont {Wettig}},\ }\href
  {https://doi.org/10.1146/annurev.nucl.50.1.343} {\bibfield  {journal}
  {\bibinfo  {journal} {Ann. Rev. Nucl. Part. Sci.}\ }\textbf {\bibinfo
  {volume} {50}},\ \bibinfo {pages} {343} (\bibinfo {year} {2000})},\ \Eprint
  {https://arxiv.org/abs/hep-ph/0003017} {arXiv:hep-ph/0003017} \BibitemShut
  {NoStop}%
\bibitem [{\citenamefont {Akemann}(2007)}]{Akemann:2007rf}%
  \BibitemOpen
  \bibfield  {author} {\bibinfo {author} {\bibfnamefont {G.}~\bibnamefont
  {Akemann}},\ }\href {https://doi.org/10.1142/S0217751X07036154} {\bibfield
  {journal} {\bibinfo  {journal} {Int. J. Mod. Phys. A}\ }\textbf {\bibinfo
  {volume} {22}},\ \bibinfo {pages} {1077} (\bibinfo {year} {2007})},\ \Eprint
  {https://arxiv.org/abs/hep-th/0701175} {arXiv:hep-th/0701175} \BibitemShut
  {NoStop}%
\bibitem [{\citenamefont {Verbaarschot}(2015)}]{Verbaarschot:2009jz}%
  \BibitemOpen
  \bibfield  {author} {\bibinfo {author} {\bibfnamefont {J.~J.~M.}\
  \bibnamefont {Verbaarschot}},\ }in\ \href
  {https://doi.org/10.1093/oxfordhb/9780198744191.001.0001} {\emph {\bibinfo
  {booktitle} {{The Oxford Handbook of Random Matrix Theory}}}},\ \bibinfo
  {editor} {edited by\ \bibinfo {editor} {\bibfnamefont {G.}~\bibnamefont
  {Akemann}}, \bibinfo {editor} {\bibfnamefont {J.}~\bibnamefont {Baik}},\ and\
  \bibinfo {editor} {\bibfnamefont {P.}~\bibnamefont {Di~Francesco}}}\
  (\bibinfo  {publisher} {Oxford University Press},\ \bibinfo {year} {2015})\
  Chap.~\bibinfo {chapter} {32},\ \Eprint {https://arxiv.org/abs/0910.4134}
  {arXiv:0910.4134 [hep-th]} \BibitemShut {NoStop}%
\bibitem [{\citenamefont {Stephanov}(1996)}]{Stephanov:1996ki}%
  \BibitemOpen
  \bibfield  {author} {\bibinfo {author} {\bibfnamefont {M.~A.}\ \bibnamefont
  {Stephanov}},\ }\href {https://doi.org/10.1103/PhysRevLett.76.4472}
  {\bibfield  {journal} {\bibinfo  {journal} {Phys. Rev. Lett.}\ }\textbf
  {\bibinfo {volume} {76}},\ \bibinfo {pages} {4472} (\bibinfo {year}
  {1996})},\ \Eprint {https://arxiv.org/abs/hep-lat/9604003}
  {arXiv:hep-lat/9604003} \BibitemShut {NoStop}%
\bibitem [{\citenamefont {Janik}\ \emph {et~al.}(1996)\citenamefont {Janik},
  \citenamefont {Nowak}, \citenamefont {Papp},\ and\ \citenamefont
  {Zahed}}]{Janik:1996va}%
  \BibitemOpen
  \bibfield  {author} {\bibinfo {author} {\bibfnamefont {R.~A.}\ \bibnamefont
  {Janik}}, \bibinfo {author} {\bibfnamefont {M.~A.}\ \bibnamefont {Nowak}},
  \bibinfo {author} {\bibfnamefont {G.}~\bibnamefont {Papp}},\ and\ \bibinfo
  {author} {\bibfnamefont {I.}~\bibnamefont {Zahed}},\ }\href
  {https://doi.org/10.1103/PhysRevLett.77.4876} {\bibfield  {journal} {\bibinfo
   {journal} {Phys. Rev. Lett.}\ }\textbf {\bibinfo {volume} {77}},\ \bibinfo
  {pages} {4876} (\bibinfo {year} {1996})},\ \Eprint
  {https://arxiv.org/abs/hep-ph/9606329} {arXiv:hep-ph/9606329} \BibitemShut
  {NoStop}%
\bibitem [{\citenamefont {Hal{\'a}sz}\ \emph
  {et~al.}(1997{\natexlab{a}})\citenamefont {Hal{\'a}sz}, \citenamefont
  {Jackson},\ and\ \citenamefont {Verbaarschot}}]{Halasz:1996jg}%
  \BibitemOpen
  \bibfield  {author} {\bibinfo {author} {\bibfnamefont {{\'A}.~M.}\
  \bibnamefont {Hal{\'a}sz}}, \bibinfo {author} {\bibfnamefont {A.~D.}\
  \bibnamefont {Jackson}},\ and\ \bibinfo {author} {\bibfnamefont {J.~J.~M.}\
  \bibnamefont {Verbaarschot}},\ }\href
  {https://doi.org/10.1016/S0370-2693(97)00015-4} {\bibfield  {journal}
  {\bibinfo  {journal} {Phys. Lett. B}\ }\textbf {\bibinfo {volume} {395}},\
  \bibinfo {pages} {293} (\bibinfo {year} {1997}{\natexlab{a}})},\ \Eprint
  {https://arxiv.org/abs/hep-lat/9611008} {arXiv:hep-lat/9611008} \BibitemShut
  {NoStop}%
\bibitem [{\citenamefont {Hal{\'a}sz}\ \emph
  {et~al.}(1997{\natexlab{b}})\citenamefont {Hal{\'a}sz}, \citenamefont
  {Jackson},\ and\ \citenamefont {Verbaarschot}}]{Halasz:1997he}%
  \BibitemOpen
  \bibfield  {author} {\bibinfo {author} {\bibfnamefont {{\'A}.~M.}\
  \bibnamefont {Hal{\'a}sz}}, \bibinfo {author} {\bibfnamefont {A.~D.}\
  \bibnamefont {Jackson}},\ and\ \bibinfo {author} {\bibfnamefont {J.~J.~M.}\
  \bibnamefont {Verbaarschot}},\ }\href
  {https://doi.org/10.1103/PhysRevD.56.5140} {\bibfield  {journal} {\bibinfo
  {journal} {Phys. Rev. D}\ }\textbf {\bibinfo {volume} {56}},\ \bibinfo
  {pages} {5140} (\bibinfo {year} {1997}{\natexlab{b}})},\ \Eprint
  {https://arxiv.org/abs/hep-lat/9703006} {arXiv:hep-lat/9703006} \BibitemShut
  {NoStop}%
\bibitem [{\citenamefont {Feinberg}\ and\ \citenamefont
  {Zee}(1997)}]{Feinberg:1997dk}%
  \BibitemOpen
  \bibfield  {author} {\bibinfo {author} {\bibfnamefont {J.}~\bibnamefont
  {Feinberg}}\ and\ \bibinfo {author} {\bibfnamefont {A.}~\bibnamefont {Zee}},\
  }\href {https://doi.org/10.1016/S0550-3213(97)00502-6} {\bibfield  {journal}
  {\bibinfo  {journal} {Nucl. Phys. B}\ }\textbf {\bibinfo {volume} {504}},\
  \bibinfo {pages} {579} (\bibinfo {year} {1997})},\ \Eprint
  {https://arxiv.org/abs/cond-mat/9703087} {arXiv:cond-mat/9703087}
  \BibitemShut {NoStop}%
\bibitem [{\citenamefont {Hal{\'a}sz}\ \emph {et~al.}(1998)\citenamefont
  {Hal{\'a}sz}, \citenamefont {Jackson}, \citenamefont {Shrock}, \citenamefont
  {Stephanov},\ and\ \citenamefont {Verbaarschot}}]{Halasz:1998qr}%
  \BibitemOpen
  \bibfield  {author} {\bibinfo {author} {\bibfnamefont {{\'A}.~M.}\
  \bibnamefont {Hal{\'a}sz}}, \bibinfo {author} {\bibfnamefont {A.~D.}\
  \bibnamefont {Jackson}}, \bibinfo {author} {\bibfnamefont {R.~E.}\
  \bibnamefont {Shrock}}, \bibinfo {author} {\bibfnamefont {M.~A.}\
  \bibnamefont {Stephanov}},\ and\ \bibinfo {author} {\bibfnamefont {J.~J.~M.}\
  \bibnamefont {Verbaarschot}},\ }\href
  {https://doi.org/10.1103/PhysRevD.58.096007} {\bibfield  {journal} {\bibinfo
  {journal} {Phys. Rev. D}\ }\textbf {\bibinfo {volume} {58}},\ \bibinfo
  {pages} {096007} (\bibinfo {year} {1998})},\ \Eprint
  {https://arxiv.org/abs/hep-ph/9804290} {arXiv:hep-ph/9804290} \BibitemShut
  {NoStop}%
\bibitem [{\citenamefont {Akemann}(2002{\natexlab{a}})}]{Akemann:2002ym}%
  \BibitemOpen
  \bibfield  {author} {\bibinfo {author} {\bibfnamefont {G.}~\bibnamefont
  {Akemann}},\ }\href {https://doi.org/10.1103/PhysRevLett.89.072002}
  {\bibfield  {journal} {\bibinfo  {journal} {Phys. Rev. Lett.}\ }\textbf
  {\bibinfo {volume} {89}},\ \bibinfo {pages} {072002} (\bibinfo {year}
  {2002}{\natexlab{a}})},\ \Eprint {https://arxiv.org/abs/hep-th/0204068}
  {arXiv:hep-th/0204068} \BibitemShut {NoStop}%
\bibitem [{\citenamefont {Klein}\ \emph {et~al.}(2003)\citenamefont {Klein},
  \citenamefont {Toublan},\ and\ \citenamefont {Verbaarschot}}]{Klein:2003fy}%
  \BibitemOpen
  \bibfield  {author} {\bibinfo {author} {\bibfnamefont {B.}~\bibnamefont
  {Klein}}, \bibinfo {author} {\bibfnamefont {D.}~\bibnamefont {Toublan}},\
  and\ \bibinfo {author} {\bibfnamefont {J.~J.~M.}\ \bibnamefont
  {Verbaarschot}},\ }\href {https://doi.org/10.1103/PhysRevD.68.014009}
  {\bibfield  {journal} {\bibinfo  {journal} {Phys. Rev. D}\ }\textbf {\bibinfo
  {volume} {68}},\ \bibinfo {pages} {014009} (\bibinfo {year} {2003})},\
  \Eprint {https://arxiv.org/abs/hep-ph/0301143} {arXiv:hep-ph/0301143}
  \BibitemShut {NoStop}%
\bibitem [{\citenamefont {Akemann}\ and\ \citenamefont
  {Wettig}(2004)}]{Akemann:2003wg}%
  \BibitemOpen
  \bibfield  {author} {\bibinfo {author} {\bibfnamefont {G.}~\bibnamefont
  {Akemann}}\ and\ \bibinfo {author} {\bibfnamefont {T.}~\bibnamefont
  {Wettig}},\ }\href {https://doi.org/10.1103/PhysRevLett.92.102002} {\bibfield
   {journal} {\bibinfo  {journal} {Phys. Rev. Lett.}\ }\textbf {\bibinfo
  {volume} {92}},\ \bibinfo {pages} {102002} (\bibinfo {year} {2004})},\
  \bibinfo {note} {[Erratum: Phys.\ Rev.\ Lett. 96, 029902 (2006)]},\ \Eprint
  {https://arxiv.org/abs/hep-lat/0308003} {arXiv:hep-lat/0308003} \BibitemShut
  {NoStop}%
\bibitem [{\citenamefont {Osborn}(2004)}]{Osborn:2004rf}%
  \BibitemOpen
  \bibfield  {author} {\bibinfo {author} {\bibfnamefont {J.~C.}\ \bibnamefont
  {Osborn}},\ }\href {https://doi.org/10.1103/PhysRevLett.93.222001} {\bibfield
   {journal} {\bibinfo  {journal} {Phys. Rev. Lett.}\ }\textbf {\bibinfo
  {volume} {93}},\ \bibinfo {pages} {222001} (\bibinfo {year} {2004})},\
  \Eprint {https://arxiv.org/abs/hep-th/0403131} {arXiv:hep-th/0403131}
  \BibitemShut {NoStop}%
\bibitem [{\citenamefont {Akemann}\ \emph {et~al.}(2005)\citenamefont
  {Akemann}, \citenamefont {Osborn}, \citenamefont {Splittorff},\ and\
  \citenamefont {Verbaarschot}}]{Akemann:2004dr}%
  \BibitemOpen
  \bibfield  {author} {\bibinfo {author} {\bibfnamefont {G.}~\bibnamefont
  {Akemann}}, \bibinfo {author} {\bibfnamefont {J.~C.}\ \bibnamefont {Osborn}},
  \bibinfo {author} {\bibfnamefont {K.}~\bibnamefont {Splittorff}},\ and\
  \bibinfo {author} {\bibfnamefont {J.~J.~M.}\ \bibnamefont {Verbaarschot}},\
  }\href {https://doi.org/10.1016/j.nuclphysb.2005.01.018} {\bibfield
  {journal} {\bibinfo  {journal} {Nucl. Phys. B}\ }\textbf {\bibinfo {volume}
  {712}},\ \bibinfo {pages} {287} (\bibinfo {year} {2005})},\ \Eprint
  {https://arxiv.org/abs/hep-th/0411030} {arXiv:hep-th/0411030} \BibitemShut
  {NoStop}%
\bibitem [{\citenamefont {Jackson}\ and\ \citenamefont
  {Verbaarschot}(1996)}]{Jackson:1995nf}%
  \BibitemOpen
  \bibfield  {author} {\bibinfo {author} {\bibfnamefont {A.~D.}\ \bibnamefont
  {Jackson}}\ and\ \bibinfo {author} {\bibfnamefont {J.~J.~M.}\ \bibnamefont
  {Verbaarschot}},\ }\href {https://doi.org/10.1103/PhysRevD.53.7223}
  {\bibfield  {journal} {\bibinfo  {journal} {Phys. Rev. D}\ }\textbf {\bibinfo
  {volume} {53}},\ \bibinfo {pages} {7223} (\bibinfo {year} {1996})},\ \Eprint
  {https://arxiv.org/abs/hep-ph/9509324} {arXiv:hep-ph/9509324} \BibitemShut
  {NoStop}%
\bibitem [{\citenamefont {Wettig}\ \emph {et~al.}(1996)\citenamefont {Wettig},
  \citenamefont {Sch{\"a}fer},\ and\ \citenamefont
  {Weidenm{\"u}ller}}]{Wettig:1995fg}%
  \BibitemOpen
  \bibfield  {author} {\bibinfo {author} {\bibfnamefont {T.}~\bibnamefont
  {Wettig}}, \bibinfo {author} {\bibfnamefont {A.}~\bibnamefont
  {Sch{\"a}fer}},\ and\ \bibinfo {author} {\bibfnamefont {H.~A.}\ \bibnamefont
  {Weidenm{\"u}ller}},\ }\href {https://doi.org/10.1016/0370-2693(95)01401-2}
  {\bibfield  {journal} {\bibinfo  {journal} {Phys. Lett. B}\ }\textbf
  {\bibinfo {volume} {367}},\ \bibinfo {pages} {28} (\bibinfo {year} {1996})},\
  \bibinfo {note} {[Erratum: Phys.\ Lett.\ B 374, 362 (1996)]},\ \Eprint
  {https://arxiv.org/abs/hep-ph/9510258} {arXiv:hep-ph/9510258} \BibitemShut
  {NoStop}%
\bibitem [{\citenamefont {Jackson}\ \emph {et~al.}(1996)\citenamefont
  {Jackson}, \citenamefont {{\c S}ener},\ and\ \citenamefont
  {Verbaarschot}}]{Jackson:1996xt}%
  \BibitemOpen
  \bibfield  {author} {\bibinfo {author} {\bibfnamefont {A.~D.}\ \bibnamefont
  {Jackson}}, \bibinfo {author} {\bibfnamefont {M.~K.}\ \bibnamefont {{\c
  S}ener}},\ and\ \bibinfo {author} {\bibfnamefont {J.~J.~M.}\ \bibnamefont
  {Verbaarschot}},\ }\href {https://doi.org/10.1016/0550-3213(96)00397-5}
  {\bibfield  {journal} {\bibinfo  {journal} {Nucl. Phys. B}\ }\textbf
  {\bibinfo {volume} {479}},\ \bibinfo {pages} {707} (\bibinfo {year}
  {1996})},\ \Eprint {https://arxiv.org/abs/hep-ph/9602225}
  {arXiv:hep-ph/9602225} \BibitemShut {NoStop}%
\bibitem [{\citenamefont {Han}\ and\ \citenamefont
  {Stephanov}(2008)}]{Han:2008xj}%
  \BibitemOpen
  \bibfield  {author} {\bibinfo {author} {\bibfnamefont {J.}~\bibnamefont
  {Han}}\ and\ \bibinfo {author} {\bibfnamefont {M.~A.}\ \bibnamefont
  {Stephanov}},\ }\href {https://doi.org/10.1103/PhysRevD.78.054507} {\bibfield
   {journal} {\bibinfo  {journal} {Phys. Rev. D}\ }\textbf {\bibinfo {volume}
  {78}},\ \bibinfo {pages} {054507} (\bibinfo {year} {2008})},\ \Eprint
  {https://arxiv.org/abs/0805.1939} {arXiv:0805.1939 [hep-lat]} \BibitemShut
  {NoStop}%
\bibitem [{\citenamefont {Bloch}\ \emph {et~al.}(2018)\citenamefont {Bloch},
  \citenamefont {Glesaaen}, \citenamefont {Verbaarschot},\ and\ \citenamefont
  {Zafeiropoulos}}]{Bloch:2017sex}%
  \BibitemOpen
  \bibfield  {author} {\bibinfo {author} {\bibfnamefont {J.}~\bibnamefont
  {Bloch}}, \bibinfo {author} {\bibfnamefont {J.}~\bibnamefont {Glesaaen}},
  \bibinfo {author} {\bibfnamefont {J.~J.~M.}\ \bibnamefont {Verbaarschot}},\
  and\ \bibinfo {author} {\bibfnamefont {S.}~\bibnamefont {Zafeiropoulos}},\
  }\href {https://doi.org/10.1007/JHEP03(2018)015} {\bibfield  {journal}
  {\bibinfo  {journal} {J. High Energy Phys.}\ }\textbf {\bibinfo {volume}
  {03}}\bibfield  {number} {\bibinfo  {number} { (2018)},\ \bibinfo {pages}
  {015}},\ }\Eprint {https://arxiv.org/abs/1712.07514} {arXiv:1712.07514
  [hep-lat]} \BibitemShut {NoStop}%
\bibitem [{\citenamefont {Giordano}\ \emph {et~al.}(2023)\citenamefont
  {Giordano}, \citenamefont {P{\'a}sztor}, \citenamefont {Peszny{\'a}k},\ and\
  \citenamefont {Tulip{\'a}nt}}]{Giordano:2023ppk}%
  \BibitemOpen
  \bibfield  {author} {\bibinfo {author} {\bibfnamefont {M.}~\bibnamefont
  {Giordano}}, \bibinfo {author} {\bibfnamefont {A.}~\bibnamefont
  {P{\'a}sztor}}, \bibinfo {author} {\bibfnamefont {D.}~\bibnamefont
  {Peszny{\'a}k}},\ and\ \bibinfo {author} {\bibfnamefont {Z.}~\bibnamefont
  {Tulip{\'a}nt}},\ }\href {https://doi.org/10.1103/PhysRevD.108.094507}
  {\bibfield  {journal} {\bibinfo  {journal} {Phys. Rev. D}\ }\textbf {\bibinfo
  {volume} {108}},\ \bibinfo {pages} {094507} (\bibinfo {year} {2023})},\
  \Eprint {https://arxiv.org/abs/2301.12947} {arXiv:2301.12947 [hep-lat]}
  \BibitemShut {NoStop}%
\bibitem [{\citenamefont {Vanderheyden}\ and\ \citenamefont
  {Jackson}(2000)}]{Vanderheyden:2000ti}%
  \BibitemOpen
  \bibfield  {author} {\bibinfo {author} {\bibfnamefont {B.}~\bibnamefont
  {Vanderheyden}}\ and\ \bibinfo {author} {\bibfnamefont {A.~D.}\ \bibnamefont
  {Jackson}},\ }\href {https://doi.org/10.1103/PhysRevD.62.094010} {\bibfield
  {journal} {\bibinfo  {journal} {Phys. Rev. D}\ }\textbf {\bibinfo {volume}
  {62}},\ \bibinfo {pages} {094010} (\bibinfo {year} {2000})},\ \Eprint
  {https://arxiv.org/abs/hep-ph/0003150} {arXiv:hep-ph/0003150} \BibitemShut
  {NoStop}%
\bibitem [{\citenamefont {Vanderheyden}\ and\ \citenamefont
  {Jackson}(2011)}]{Vanderheyden:2011iq}%
  \BibitemOpen
  \bibfield  {author} {\bibinfo {author} {\bibfnamefont {B.}~\bibnamefont
  {Vanderheyden}}\ and\ \bibinfo {author} {\bibfnamefont {A.~D.}\ \bibnamefont
  {Jackson}},\ }\href {https://doi.org/10.1088/0034-4885/74/10/102001}
  {\bibfield  {journal} {\bibinfo  {journal} {Rept. Prog. Phys.}\ }\textbf
  {\bibinfo {volume} {74}},\ \bibinfo {pages} {102001} (\bibinfo {year}
  {2011})},\ \Eprint {https://arxiv.org/abs/1105.1291} {arXiv:1105.1291
  [hep-ph]} \BibitemShut {NoStop}%
\bibitem [{\citenamefont {Ambj{\o}rn}\ \emph {et~al.}(1992)\citenamefont
  {Ambj{\o}rn}, \citenamefont {Kristjansen},\ and\ \citenamefont
  {Makeenko}}]{Ambjorn:1992xu}%
  \BibitemOpen
  \bibfield  {author} {\bibinfo {author} {\bibfnamefont {J.}~\bibnamefont
  {Ambj{\o}rn}}, \bibinfo {author} {\bibfnamefont {C.~F.}\ \bibnamefont
  {Kristjansen}},\ and\ \bibinfo {author} {\bibfnamefont {Y.~M.}\ \bibnamefont
  {Makeenko}},\ }\href {https://doi.org/10.1142/S0217732392002573} {\bibfield
  {journal} {\bibinfo  {journal} {Mod. Phys. Lett. A}\ }\textbf {\bibinfo
  {volume} {7}},\ \bibinfo {pages} {3187} (\bibinfo {year} {1992})},\ \Eprint
  {https://arxiv.org/abs/hep-th/9207020} {arXiv:hep-th/9207020} \BibitemShut
  {NoStop}%
\bibitem [{\citenamefont {Kanzieper}\ and\ \citenamefont
  {Freilikher}(1999)}]{Kanzieper:1998ti}%
  \BibitemOpen
  \bibfield  {author} {\bibinfo {author} {\bibfnamefont {E.}~\bibnamefont
  {Kanzieper}}\ and\ \bibinfo {author} {\bibfnamefont {V.}~\bibnamefont
  {Freilikher}},\ }\href@noop {} {\bibfield  {journal} {\bibinfo  {journal}
  {NATO Sci. Ser. C}\ }\textbf {\bibinfo {volume} {531}},\ \bibinfo {pages}
  {165} (\bibinfo {year} {1999})},\ \Eprint
  {https://arxiv.org/abs/cond-mat/9809365} {arXiv:cond-mat/9809365}
  \BibitemShut {NoStop}%
\bibitem [{\citenamefont {Akemann}(2002{\natexlab{b}})}]{Akemann:2002ch}%
  \BibitemOpen
  \bibfield  {author} {\bibinfo {author} {\bibfnamefont {G.}~\bibnamefont
  {Akemann}},\ }\href {https://doi.org/10.1016/S0370-2693(02)02737-5}
  {\bibfield  {journal} {\bibinfo  {journal} {Phys. Lett. B}\ }\textbf
  {\bibinfo {volume} {547}},\ \bibinfo {pages} {100} (\bibinfo {year}
  {2002}{\natexlab{b}})},\ \Eprint {https://arxiv.org/abs/hep-th/0206086}
  {arXiv:hep-th/0206086} \BibitemShut {NoStop}%
\bibitem [{\citenamefont {Witten}(2011)}]{Witten:2010cx}%
  \BibitemOpen
  \bibfield  {author} {\bibinfo {author} {\bibfnamefont {E.}~\bibnamefont
  {Witten}},\ }\href@noop {} {\bibfield  {journal} {\bibinfo  {journal} {AMS/IP
  Stud. Adv. Math.}\ }\textbf {\bibinfo {volume} {50}},\ \bibinfo {pages} {347}
  (\bibinfo {year} {2011})},\ \Eprint {https://arxiv.org/abs/1001.2933}
  {arXiv:1001.2933 [hep-th]} \BibitemShut {NoStop}%
\bibitem [{\citenamefont {Cristoforetti}\ \emph {et~al.}(2012)\citenamefont
  {Cristoforetti}, \citenamefont {Di~Renzo},\ and\ \citenamefont
  {Scorzato}}]{Cristoforetti:2012su}%
  \BibitemOpen
  \bibfield  {author} {\bibinfo {author} {\bibfnamefont {M.}~\bibnamefont
  {Cristoforetti}}, \bibinfo {author} {\bibfnamefont {F.}~\bibnamefont
  {Di~Renzo}},\ and\ \bibinfo {author} {\bibfnamefont {L.}~\bibnamefont
  {Scorzato}} (\bibinfo {collaboration} {AuroraScience}),\ }\href
  {https://doi.org/10.1103/PhysRevD.86.074506} {\bibfield  {journal} {\bibinfo
  {journal} {Phys. Rev. D}\ }\textbf {\bibinfo {volume} {86}},\ \bibinfo
  {pages} {074506} (\bibinfo {year} {2012})},\ \Eprint
  {https://arxiv.org/abs/1205.3996} {arXiv:1205.3996 [hep-lat]} \BibitemShut
  {NoStop}%
\bibitem [{\citenamefont {Vafa}\ and\ \citenamefont
  {Witten}(1984)}]{Vafa:1983tf}%
  \BibitemOpen
  \bibfield  {author} {\bibinfo {author} {\bibfnamefont {C.}~\bibnamefont
  {Vafa}}\ and\ \bibinfo {author} {\bibfnamefont {E.}~\bibnamefont {Witten}},\
  }\href {https://doi.org/10.1016/0550-3213(84)90230-X} {\bibfield  {journal}
  {\bibinfo  {journal} {Nucl. Phys. B}\ }\textbf {\bibinfo {volume} {234}},\
  \bibinfo {pages} {173} (\bibinfo {year} {1984})}\BibitemShut {NoStop}%
\bibitem [{\citenamefont {Aloisio}\ \emph {et~al.}(2001)\citenamefont
  {Aloisio}, \citenamefont {Azcoiti}, \citenamefont {Di~Carlo}, \citenamefont
  {Galante},\ and\ \citenamefont {Grillo}}]{Aloisio:2000rb}%
  \BibitemOpen
  \bibfield  {author} {\bibinfo {author} {\bibfnamefont {R.}~\bibnamefont
  {Aloisio}}, \bibinfo {author} {\bibfnamefont {V.}~\bibnamefont {Azcoiti}},
  \bibinfo {author} {\bibfnamefont {G.}~\bibnamefont {Di~Carlo}}, \bibinfo
  {author} {\bibfnamefont {A.}~\bibnamefont {Galante}},\ and\ \bibinfo {author}
  {\bibfnamefont {A.~F.}\ \bibnamefont {Grillo}},\ }\href
  {https://doi.org/10.1016/S0550-3213(01)00232-2} {\bibfield  {journal}
  {\bibinfo  {journal} {Nucl. Phys. B}\ }\textbf {\bibinfo {volume} {606}},\
  \bibinfo {pages} {322} (\bibinfo {year} {2001})},\ \Eprint
  {https://arxiv.org/abs/hep-lat/0011079} {arXiv:hep-lat/0011079} \BibitemShut
  {NoStop}%
\bibitem [{\citenamefont {Giordano}(2023)}]{Giordano:2023spj}%
  \BibitemOpen
  \bibfield  {author} {\bibinfo {author} {\bibfnamefont {M.}~\bibnamefont
  {Giordano}},\ }\href {https://doi.org/10.1103/PhysRevD.107.114509} {\bibfield
   {journal} {\bibinfo  {journal} {Phys. Rev. D}\ }\textbf {\bibinfo {volume}
  {107}},\ \bibinfo {pages} {114509} (\bibinfo {year} {2023})},\ \Eprint
  {https://arxiv.org/abs/2303.03109} {arXiv:2303.03109 [hep-lat]} \BibitemShut
  {NoStop}%
\bibitem [{\citenamefont {Lucini}\ \emph {et~al.}(2007)\citenamefont {Lucini},
  \citenamefont {Patella},\ and\ \citenamefont {Pica}}]{Lucini:2007im}%
  \BibitemOpen
  \bibfield  {author} {\bibinfo {author} {\bibfnamefont {B.}~\bibnamefont
  {Lucini}}, \bibinfo {author} {\bibfnamefont {A.}~\bibnamefont {Patella}},\
  and\ \bibinfo {author} {\bibfnamefont {C.}~\bibnamefont {Pica}},\ }\href
  {https://doi.org/10.1063/1.2823768} {\bibfield  {journal} {\bibinfo
  {journal} {AIP Conf. Proc.}\ }\textbf {\bibinfo {volume} {957}},\ \bibinfo
  {pages} {229} (\bibinfo {year} {2007})},\ \Eprint
  {https://arxiv.org/abs/0709.0909} {arXiv:0709.0909 [hep-lat]} \BibitemShut
  {NoStop}%
\bibitem [{\citenamefont {Lucini}\ and\ \citenamefont
  {Patella}(2009)}]{Lucini:2009kf}%
  \BibitemOpen
  \bibfield  {author} {\bibinfo {author} {\bibfnamefont {B.}~\bibnamefont
  {Lucini}}\ and\ \bibinfo {author} {\bibfnamefont {A.}~\bibnamefont
  {Patella}},\ }\href {https://doi.org/10.1103/PhysRevD.79.125030} {\bibfield
  {journal} {\bibinfo  {journal} {Phys. Rev. D}\ }\textbf {\bibinfo {volume}
  {79}},\ \bibinfo {pages} {125030} (\bibinfo {year} {2009})},\ \Eprint
  {https://arxiv.org/abs/0904.3479} {arXiv:0904.3479 [hep-th]} \BibitemShut
  {NoStop}%
\bibitem [{\citenamefont {Alexandru}\ \emph {et~al.}(2018)\citenamefont
  {Alexandru}, \citenamefont {Bedaque}, \citenamefont {Lamm},\ and\
  \citenamefont {Lawrence}}]{Alexandru:2018fqp}%
  \BibitemOpen
  \bibfield  {author} {\bibinfo {author} {\bibfnamefont {A.}~\bibnamefont
  {Alexandru}}, \bibinfo {author} {\bibfnamefont {P.~F.}\ \bibnamefont
  {Bedaque}}, \bibinfo {author} {\bibfnamefont {H.}~\bibnamefont {Lamm}},\ and\
  \bibinfo {author} {\bibfnamefont {S.}~\bibnamefont {Lawrence}},\ }\href
  {https://doi.org/10.1103/PhysRevD.97.094510} {\bibfield  {journal} {\bibinfo
  {journal} {Phys. Rev. D}\ }\textbf {\bibinfo {volume} {97}},\ \bibinfo
  {pages} {094510} (\bibinfo {year} {2018})},\ \Eprint
  {https://arxiv.org/abs/1804.00697} {arXiv:1804.00697 [hep-lat]} \BibitemShut
  {NoStop}%
\bibitem [{\citenamefont {Giordano}\ \emph {et~al.}(2022)\citenamefont
  {Giordano}, \citenamefont {Kap{\'a}s}, \citenamefont {Katz}, \citenamefont
  {P{\'a}sztor},\ and\ \citenamefont {Tulip{\'a}nt}}]{Tulipant:2022vtk}%
  \BibitemOpen
  \bibfield  {author} {\bibinfo {author} {\bibfnamefont {M.}~\bibnamefont
  {Giordano}}, \bibinfo {author} {\bibfnamefont {K.}~\bibnamefont {Kap{\'a}s}},
  \bibinfo {author} {\bibfnamefont {S.~D.}\ \bibnamefont {Katz}}, \bibinfo
  {author} {\bibfnamefont {A.}~\bibnamefont {P{\'a}sztor}},\ and\ \bibinfo
  {author} {\bibfnamefont {Z.}~\bibnamefont {Tulip{\'a}nt}},\ }\href
  {https://doi.org/10.22323/1.430.0161} {\bibfield  {journal} {\bibinfo
  {journal} {Phys. Rev. D}\ }\textbf {\bibinfo {volume} {106}},\ \bibinfo
  {pages} {054512} (\bibinfo {year} {2022})},\ \Eprint
  {https://arxiv.org/abs/2202.07561} {arXiv:2202.07561 [hep-lat]} \BibitemShut
  {NoStop}%
\bibitem [{\citenamefont {Rodekamp}\ \emph {et~al.}(2022)\citenamefont
  {Rodekamp}, \citenamefont {Berkowitz}, \citenamefont {G\"antgen},
  \citenamefont {Krieg}, \citenamefont {Luu},\ and\ \citenamefont
  {Ostmeyer}}]{Rodekamp:2022xpf}%
  \BibitemOpen
  \bibfield  {author} {\bibinfo {author} {\bibfnamefont {M.}~\bibnamefont
  {Rodekamp}}, \bibinfo {author} {\bibfnamefont {E.}~\bibnamefont {Berkowitz}},
  \bibinfo {author} {\bibfnamefont {C.}~\bibnamefont {G\"antgen}}, \bibinfo
  {author} {\bibfnamefont {S.}~\bibnamefont {Krieg}}, \bibinfo {author}
  {\bibfnamefont {T.}~\bibnamefont {Luu}},\ and\ \bibinfo {author}
  {\bibfnamefont {J.}~\bibnamefont {Ostmeyer}},\ }\href
  {https://doi.org/10.1103/PhysRevB.106.125139} {\bibfield  {journal} {\bibinfo
   {journal} {Phys. Rev. B}\ }\textbf {\bibinfo {volume} {106}},\ \bibinfo
  {pages} {125139} (\bibinfo {year} {2022})},\ \Eprint
  {https://arxiv.org/abs/2203.00390} {arXiv:2203.00390 [physics.comp-ph]}
  \BibitemShut {NoStop}%
\bibitem [{\citenamefont {G\"antgen}\ \emph {et~al.}(2024)\citenamefont
  {G\"antgen}, \citenamefont {Berkowitz}, \citenamefont {Luu}, \citenamefont
  {Ostmeyer},\ and\ \citenamefont {Rodekamp}}]{Gantgen:2023byf}%
  \BibitemOpen
  \bibfield  {author} {\bibinfo {author} {\bibfnamefont {C.}~\bibnamefont
  {G\"antgen}}, \bibinfo {author} {\bibfnamefont {E.}~\bibnamefont
  {Berkowitz}}, \bibinfo {author} {\bibfnamefont {T.}~\bibnamefont {Luu}},
  \bibinfo {author} {\bibfnamefont {J.}~\bibnamefont {Ostmeyer}},\ and\
  \bibinfo {author} {\bibfnamefont {M.}~\bibnamefont {Rodekamp}},\ }\href
  {https://doi.org/10.1103/PhysRevB.109.195158} {\bibfield  {journal} {\bibinfo
   {journal} {Phys. Rev. B}\ }\textbf {\bibinfo {volume} {109}},\ \bibinfo
  {pages} {195158} (\bibinfo {year} {2024})},\ \Eprint
  {https://arxiv.org/abs/2307.06785} {arXiv:2307.06785 [cond-mat.str-el]}
  \BibitemShut {NoStop}%
\bibitem [{\citenamefont {Cohen}(2003)}]{Cohen:2003kd}%
  \BibitemOpen
  \bibfield  {author} {\bibinfo {author} {\bibfnamefont {T.~D.}\ \bibnamefont
  {Cohen}},\ }\href {https://doi.org/10.1103/PhysRevLett.91.222001} {\bibfield
  {journal} {\bibinfo  {journal} {Phys. Rev. Lett.}\ }\textbf {\bibinfo
  {volume} {91}},\ \bibinfo {pages} {222001} (\bibinfo {year} {2003})},\
  \Eprint {https://arxiv.org/abs/hep-ph/0307089} {arXiv:hep-ph/0307089}
  \BibitemShut {NoStop}%
\bibitem [{\citenamefont {Roberge}\ and\ \citenamefont
  {Weiss}(1986)}]{Roberge:1986mm}%
  \BibitemOpen
  \bibfield  {author} {\bibinfo {author} {\bibfnamefont {A.}~\bibnamefont
  {Roberge}}\ and\ \bibinfo {author} {\bibfnamefont {N.}~\bibnamefont
  {Weiss}},\ }\href {https://doi.org/10.1016/0550-3213(86)90582-1} {\bibfield
  {journal} {\bibinfo  {journal} {Nucl. Phys. B}\ }\textbf {\bibinfo {volume}
  {275}},\ \bibinfo {pages} {734} (\bibinfo {year} {1986})}\BibitemShut
  {NoStop}%
\end{thebibliography}%
\end{document}